\documentclass[reprint,amsmath,amssymb,aps,floatfix,superscriptaddress,prx]{revtex4-2}

\usepackage{graphicx}
\usepackage{float}
\usepackage{hyperref}
\usepackage[utf8]{inputenc}
\usepackage[english]{babel}

\graphicspath{{images/}} 

\begin{document}

\title{Wavefront Shaping with a Tunable Metasurface: Creating Coldspots and Coherent Perfect Absorption at Arbitrary Frequencies}

\author {Benjamin W. Frazier}
\email{Benjamin.Frazier@jhuapl.edu}
\affiliation{Institute for Research in Electronics and Applied Physics, University of Maryland, College Park, MD 20742, USA}
\affiliation{Department of Electrical and Computer Engineering, University of Maryland, College Park, MD 20742, USA}
\affiliation{Johns Hopkins University Applied Physics Laboratory, Laurel, MD 20723, USA}

\author{Thomas M. Antonsen, Jr.}
\affiliation{Institute for Research in Electronics and Applied Physics, University of Maryland, College Park, MD 20742, USA}
\affiliation{Department of Electrical and Computer Engineering, University of Maryland, College Park, MD 20742, USA}
\affiliation{Department of Physics, University of Maryland, College Park, MD 20742, USA}

\author{Steven M. Anlage}
\affiliation{Department of Electrical and Computer Engineering, University of Maryland, College Park, MD 20742, USA}
\affiliation{Quantum Materials Center, University of Maryland, College Park, MD 20742, USA}
\affiliation{Department of Physics, University of Maryland, College Park, MD 20742, USA}

\author{Edward Ott}
\affiliation{Institute for Research in Electronics and Applied Physics, University of Maryland, College Park, MD 20742, USA}
\affiliation{Department of Electrical and Computer Engineering, University of Maryland, College Park, MD 20742, USA}
\affiliation{Department of Physics, University of Maryland, College Park, MD 20742, USA}
\date{\today}

\begin{abstract}
Modern electronic systems operate in complex electromagnetic environments and must handle noise and unwanted coupling. The capability to isolate or reject unwanted signals for mitigating vulnerabilities is critical in any practical application. In this work, we describe the use of a binary programmable metasurface to (i) control the spatial degrees of freedom for waves propagating inside an electromagnetic cavity and demonstrate the ability to create nulls in the transmission coefficient between selected ports; and (ii) create the conditions for coherent perfect absorption. Both objectives are performed at arbitrary frequencies. In the first case a novel and effective optimization algorithm is presented that selectively generates coldspots over a single frequency band or simultaneously over multiple frequency bands. We show that this algorithm is successful with multiple input port configurations and varying optimization bandwidths. In the second case we establish how this technique can be used to establish a multi-port coherent perfect absorption state for the cavity.

\end{abstract}

\maketitle
\section{\label{sec:intro}Introduction}
Extreme electromagnetic environments are prevalent in much of our day to day lives. While often unnoticed, this puts significant stress on the design of electronic systems that are expected to work under all conditions. Extraordinary care is taken to counter adverse effects whenever possible, but the operating environment is rarely known ahead of time. In addition, Electromagnetic Interference (EMI) takes the form of unwanted coupling between components. Some platforms, such as aircraft and spacecraft, can experience devastating consequences, resulting in severe mission degradation or even casualties \cite{leachr.d.ElectronicSystemsFailures1995}. Wave fields in electrically large enclosed areas such as passenger compartments in transportation systems show extreme sensitivity to frequency and geometric details even though these enclosures, termed chaotic cavities, are not intended to be reverberant. In such cavities the electromagnetic wave fields have specific statistical properties that depend upon a limited number of parameters \cite{hemmadyStatisticalPredictionMeasurement2012}. Among these is the fact that the wave field is statistically equivalent to a random superposition of plane waves \cite{berryRegularIrregularSemiclassical}. As such, we can leverage analytical tools from the active research area of quantum chaos \cite{haakeQuantumSignatures, ottChaos} in the more generalized framework of wave chaos \cite{zhengStatisticsImpedanceScattering2006b}.

Our goal is to adaptively and intelligently control fields inside complex electromagnetic environments. We show that this can be achieved through programmable metasurfaces, which increase the available Degrees of Freedom (DOF) by manipulating boundary conditions \cite{epsteinDesignApplicationsHuygens2015, chenHuygensMetasurfacesMicrowaves2018, heTunableReconfigurableMetasurfaces2019}. In recent years, these devices have enabled novel concepts in a wide variety of applications throughout the microwave and optical domains \cite{vellekoopFocusingCoherentLight2007, liElectromagneticReprogrammableCodingmetasurface2017, sabinoWavefrontSpatialphaseModulation2017, liAdaptiveFreespaceOptical2017, ameriUltraWidebandRadar2019, tangProgrammableMetasurfacebasedRF2019, imaniReviewMetasurfaceAntennas2020}. The additional DOF in turn enhance diversity in space and time \cite{lemoultManipulatingSpatiotemporalDegrees2009, kamaliAngleMultiplexedMetasurfacesEncoding2017, delhougneOptimalMultiplexingSpatially2020, grosTuningRegularCavity2020}, and allow intricate control of the underlying scattering system; combining a programmable metasurface with a cavity unlocks applications not possible with a fixed system \cite{dupreWaveFieldShapingCavities2015, delhougneSpatiotemporalWaveFront2016, jangWavefrontShapingDisorderengineered2018} and encourages research in new and under-explored domains \cite{zhaoMetasurfaceassistedMassiveBackscatter2020}. 

One such unexplored area is Coherent Perfection Absorption (CPA), where coherent excitation of a lossy system can result in complete absorption of all incident waves \cite{chongCoherentPerfectAbsorbers2010, pichlerRandomAntilasingCoherent2019}. It has applications in highly efficient notch filtering, energy conversion, and even detection; since the CPA state is extremely sensitive to parameter variation, it can be used to identify small changes in a complex scattering system \cite{chenPerfectAbsorptionComplex2020}. The ability of a metasurface to manipulate additional DOF presents a novel capability for realizing CPA states.

\begin{figure*}
\includegraphics[width = 17.2 cm]{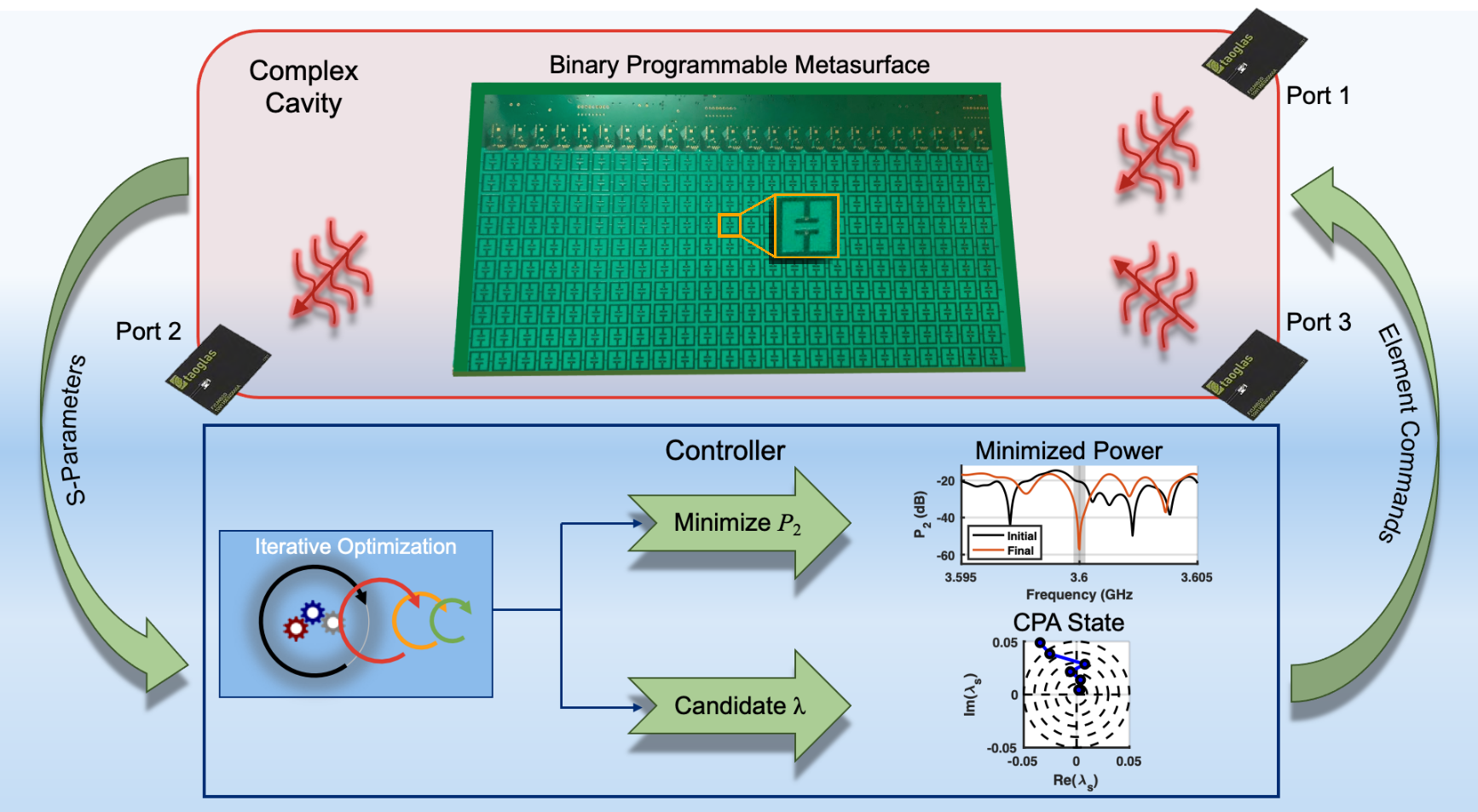}
\caption{\label{fig:overview} \bf Conceptual overview of the metasurface enabled cavity as a closed loop system. \normalfont The cavity $S$-parameters (scattering parameters) are measured with a network analyzer and passed to a controller that updates the metasurface elements  with a new set of commands. The controller can generate coldspots at port 2 at an arbitrary set of frequencies, or drive candidate $S$-matrix eigenvalues towards the origin, and includes a stochastic iterative optimization algorithm. The three ports allow additional angular and spatial diversity to be added at the inputs. The inset shows a closeup view of one of the metasurface unit cells.}  
\end{figure*}

In this article we describe the use of a binary programmable metasurface to create microwave coldspots at arbitrary frequencies, and to realize CPA states, both within the 1 GHz band of operation of the metasurface. The conceptual overview is shown in Figure \ref{fig:overview} where the metasurface is installed in a complex reverberating cavity and controlled in a closed loop manner. Input directional diversity is introduced by simultaneously driving multiple ports with arbitrary relative phase shifts. An iterative optimization algorithm is used to generate coldspots at the output port, or to drive candidate scattering matrix eigenvalues towards the origin to achieve CPA.

\section{\label{sec:configuration}Cavity Configuration}
The metasurface used for this work is a reflectarray fabricated by the Johns Hopkins University Applied Physics Laboratory (JHU/APL) that is designed to operate from 3-3.75 GHz. It has a lattice of $10 \times 24$ squares occupied by unit cells with size $< \lambda/6$ \cite{schmidSbandGaAsFET2020}, where $\lambda$ is the wavelength. These 240 unit cells can be independently set to one of two states, which approximate electric or magnetic boundary conditions and provide a relative 180$^\circ$ phase shift for waves reflected by the element. This results in the local surface impedance  of the array varying from cell to cell and state to state.  The array surface thus has $2^{240}$ independent states, each of which reflects waves in a uniquely different set of directions.  We refer to the settings of the 240 elements as a command. 

\begin{figure*}
\includegraphics[width = 17.2 cm]{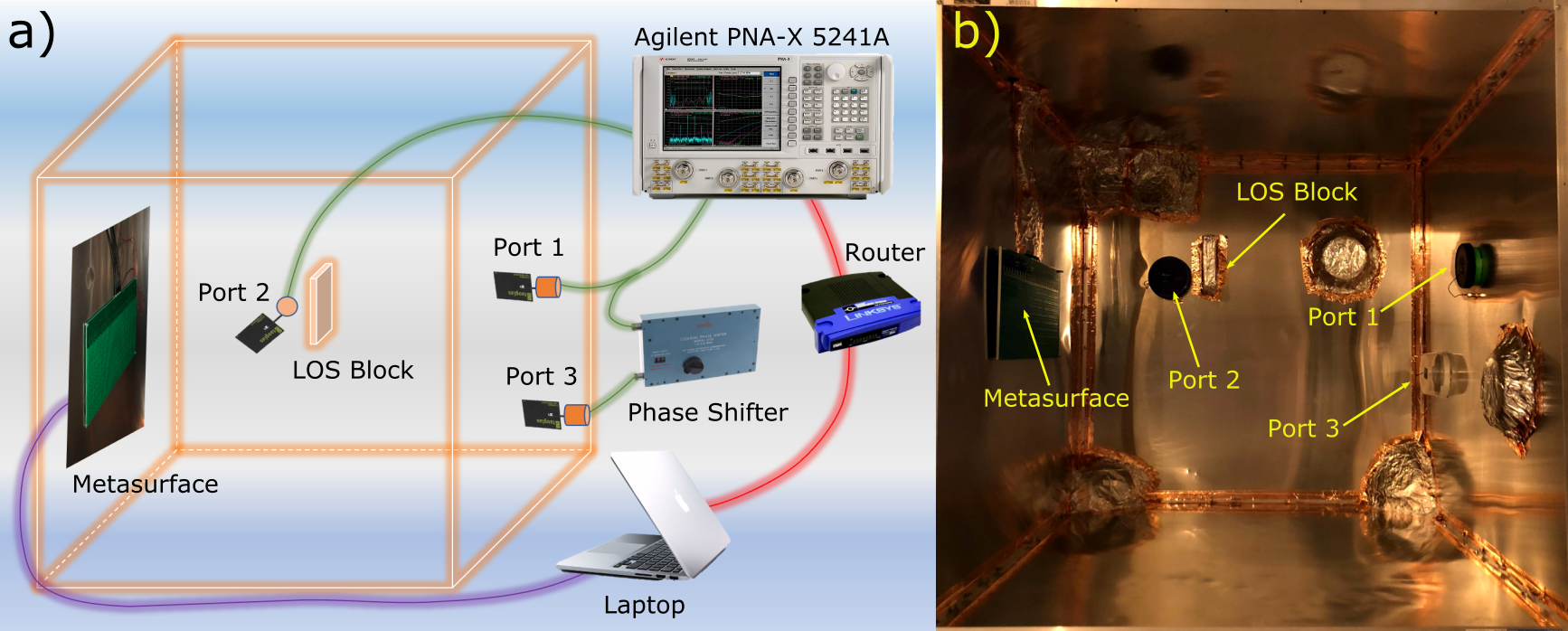}
\caption{\label{fig:geometry} \bf Experimental schematic and cavity configuration. a) \normalfont Schematic of the experimental setup in the configuration driving all 3 ports. A network analyzer (Agilent PNA-X 5241A) is used to measure cavity $S$-parameters, with channel 1 connected to both Port 1 and Port 3 (through a phase shifter) and channel 2 connected to Port 2. The ports are terminated with ultra wideband antennas. The metasurface is mounted on the cavity wall opposite Ports 1 and 3, and a block is used next to Port 2 to break the line of sight (LOS) between Port 2 and Ports 1 and 3. A laptop controls the system and is connected to the metasurface through a USB interface and to the network analyzer through a wired ethernet link. \bf b) \normalfont Photograph looking inside the cavity, with the metasurface and ports labeled. Also shown in the photo are the line of sight block and the irregular scattering elements installed on the cavity walls as discussed in the supplemental material.}  
\end{figure*}

As shown in Figure \ref{fig:geometry}, the array was installed in a 0.76 m$^3$ cavity where it covers $\sim1.5 \%$ of the total interior surface area. The cavity has 3 ports with one acting as a target for scoring and two used for signal injection; the input ports can be driven either individually or collectively with a relative phase shift. Although 3 ports are present, we are typically using the cavity as a 2-port system because we have a 2-port network analyzer. All 3 ports are used when driving ports 1 and 3 simultaneously, in which case the underlying scattering system is represented by a $3\times 3$ scattering matrix. While the experiment has a physically fixed number of ports, the results can be generalized to an arbitrary number of ports. The cavity has both low and high loss configurations to test how the behavior varies with the typical quality factor, $Q$, of the modes; here we consider the high $Q$ case. Introducing the metasurface to the cavity reduced the average $Q$ in the frequency band of operation of the metasurface by a factor of $\sim$2. However, once the metasurface was installed, the average $Q$ was found to be independent of the number of active or inactive elements on the surface. The quality factor was determined to be roughly $5.5 \times 10^3$, by measuring the power decay time, $\tau = Q/\omega$ (250 ns with the metasurface installed). Further details of the metasurface, cavity construction, experimental setup, and impact of the metasurface on losses are provided in the supplemental material.

The cavity mean mode spacing in frequency, $\Delta f$, is found from the Weyl formula as $\Delta f = \pi c^3\left(2\omega^2 V \right)^{-1}$ \cite{addissieExtractionCouplingImpedance2019}. A measure of the loss in the cavity is the $Q$-width of a mode normalized to the mode spacing, $\alpha = f/(2\Delta fQ)=3$ for this cavity.  For our cavity, the mean mode spacing is roughly 115 kHz at a 3.5 GHz center frequency. As discussed in the supplemental material, the mean spacing between nulls in the transmission coefficient, $|S_{21}|$, was found to be $\sim$2 MHz and the average width of the nulls was found to be $\sim$200 kHz. This indicates a transmission coefficient null contains about 2 modes. Alternatively, it corresponds to a path difference of 750 m between two interfering signals.

We are interested in the steady state response, so the average $Q$ does put a bound on the effective speed with which we can switch the cavity scattering matrix between fixed states. This can be seen through the power decay time, which measures how quickly energy in the cavity dissipates. In order to guarantee a steady state result, the metasurface commands must be toggled at a rate slower than $1/\tau$. The time between measurements must also be staggered by several $\tau$ to ensure each measurement corresponds to the desired metasurface commands. To observe transient behavior in a cavity with a 250 ns power decay time, the metasurface commands must be switched at rates greater than 4 MHz. This assumes the bandwidth of the measured phenomena is wider than the switching rate; an additional complication arises with narrow bandwidth responses that are of interest. When considering finite bandwidth nulls (200 kHz as stated above), the narrower bandwidth process will determine the bound. Our experimental setup is limited to switching rates $<$ 1 Hz, so neither limit presents a practical concern for our configuration. However, experiments with an embedded microcontroller demonstrated that the metasurface itself can be switched at rates up to 15 kHz \cite{schmidSbandGaAsFET2020}, so this may need to be considered with high speed operation in higher $Q$ cavities. 

A complex scattering system such as our cavity exhibits both universal fluctuations, which can be described by Random Matrix Theory (RMT) \cite{haakeQuantumSignatures}, and deterministic behavior arising from the system specific configuration of the ports and short orbits between the ports \cite{hemmadyUniversalImpedanceFluctuations2005,yehUniversalNonuniversalProperties2010, gradoniPredictingStatisticsWave2014}. Due to the small relative size of the metasurface, the chaotic ray paths with many multipath bounces will experience the strongest influence.  Since minimizing the power received at a port is accomplished by creating destructive interference of the ray paths, the relationship between commands and responses is quite complicated. This leads to utilizing stochastic iterative approaches, or machine learning, in place of linear deterministic methods for control. Here we consider iterative processes. A metasurface covering a larger fraction of the interior surface would likely produce stronger results \cite{kainaShapingComplexMicrowave2015} and allow us to use a transmission matrix based approach to determine the optimal metasurface commands \cite{popoffMeasuringTransmissionMatrix2010, dremeauReferencelessMeasurementTransmission2015, delhougneIntensityonlyMeasurementPartially2016}. For this reason most prior research utilizes metasurfaces that cover a significant portion of a wall (or multiple walls). However, using a relatively small metasurface coverage is better suited for real world applications where it is not practical to build or use a larger device.

A key step in evaluating system performance is to determine the range of possible responses of the scattering properties of the system so as to ensure that we have a sufficiently diverse command set. Unfortunately, with $2^{240}$ possible commands (approximately $1.8 \times 10^{72}$), it is not feasible to test every one and we need to find a reduced number that produces the full range of outcomes. As discussed in the supplemental material, deterministic decomposition of commands into orthogonal basis functions, such as Hadamard bases \cite{wattsTerahertzCompressiveImaging2014, wattsFrequencydivisionmultiplexedSinglepixelImaging2016}, generated a very narrow range of system scattering responses. Diversity in the responses requires a distribution of commands with a variety of spatial frequencies, ratios of active to inactive elements, and localized groupings of active elements. Doubly random methods or compound distributions, such as a biased coin toss, or power law spectral density with the bias, or power exponent itself a random draw, were found to yield the widest range of responses. Details of our novel stochastic algorithm are discussed in the next section and the supplemental material.

\section{\label{sec:optimization}Generating Coldspots}
Our goal is to program the metasurface to minimize the transmission between two ports in a complex scattering system at an arbitrary frequency. Cases are scored by evaluating the difference in average power, $\Delta P_2$, in a specified frequency range at a given center frequency between the initial inactive (all 0s) state and the current state of the metasurface. To maximize this difference we take a directed random walk approach in which at each step a number of array element states are toggled (changed), $\Delta P_2$ is evaluated, and the new state is accepted or rejected based on whether or not it decreases $\Delta P_2$. As discussed in the previous section, we need to have a mix of large and small spatial groupings of elements and a varied number of active elements to ensure a diverse set of responses. To meet this requirement, our iterative algorithm operates in 2 distinct phases: multiple element toggling and individual element toggling.

In the multiple element toggling phase, we select $M$ elements at random as a trial and toggle their state ($0\rightarrow 1$ and $1 \rightarrow 0$). If $\Delta P_2$ is decreased, the trial set of commands becomes the new reference set and we repeat the process selecting another $M$ elements at random and toggling their state. When $T$ consecutive trials have been made without improving $\Delta P_2$ we claim convergence and move to the next value of $M$. In a typical experimental run, $M = [120, 48, 24, 12, 6]$, and $T$ = 30. After all values of $M$ have been exhausted, we move to the individual element phase. 

The individual element toggling phase has 3 cases associated with each trial. We select a single element at random and toggle it and then, in an adaptation of the neighbor toggling method of Ref. \cite{dorrerDirectBinarySearch2018}, we toggle the 4 nearest neighbors and the 4 diagonal neighbors. $\Delta P_2$ is evaluated for each of these cases and the algorithm continues as in the 1st phase until $T$ consecutive trials are performed without improving $\Delta P_2$. 

The multiple element toggling phase tends to result in a local minimum which is difficult to escape when toggling only a single element. Adding neighbor toggling significantly improves the performance, as it provides larger localized changes in the command set and allows us to escape the local minimum. Even with the neighbor toggling, however, our stochastic approach does not guarantee that a global minimum is found. Increasing the convergence criteria, $T$, can increase the probability of finding the global minimum, but comes with the cost of increased time. The absolute minimum is not necessarily required, and our stochastic algorithm is able to provide substantially deep nulls at arbitrary frequencies in a reasonable amount of time. 

A typical experimental run will provide $\sim$350 trials, $\sim$25 iteration updates, and take $\sim$1.5 hours, as the experimental setup is not optimized for run time. We use an ethernet connection to transfer 32001 frequency samples over the full 1 GHz band for each of the 4 complex $S$-parameter measurements using 64-bit precision. With the frequency values themselves included, this means 2.3 MB of data are transferred for each trial. In addition, the commands and measured $|S_{21}|$ are plotted at each trial for operator feedback, resulting in a delay of $\sim$15 seconds per trial. Disabling plotting and capturing only the processed frequency band could potentially reduce the time to 1-2 seconds per trial, or 6-12 minutes for an experimental run. In general, the cavity should not be used for other purposes while generating a coldspot, as trials that move in the wrong direction in the solution space may produce undesirable responses. Reducing the time to find a solution in only a few minutes may present an acceptable interruption in service. To move towards a faster, real time operational system, we would replace the network analyzer with software defined radios (SDRs), such as the HackRF One \cite{hackrfone} or BladeRF \cite{bladerf} commercial devices which retail for $\sim$\$300 or $\sim$\$500-1000 respectively. In addition, an embedded micro-controller, such as an Arduino or Raspberry Pi, could be used to reduce the USB communications overhead induced by traditional desktop operating systems when interfacing with the metasurface. This would mean measuring signal I and Q channels rather than $S$-parameters; however, this is a realistic requirement for a practical fielded solution that would not use a bulky, expensive network analyzer anyway. Trial rates approaching 1 kHz could be achieved in this fashion, though substantial engineering effort would be required to reduce the latency to approach the metasurface switching limit of 15 kHz \cite{schmidSbandGaAsFET2020}.

\begin{figure*}
\includegraphics[width = 17.2 cm]{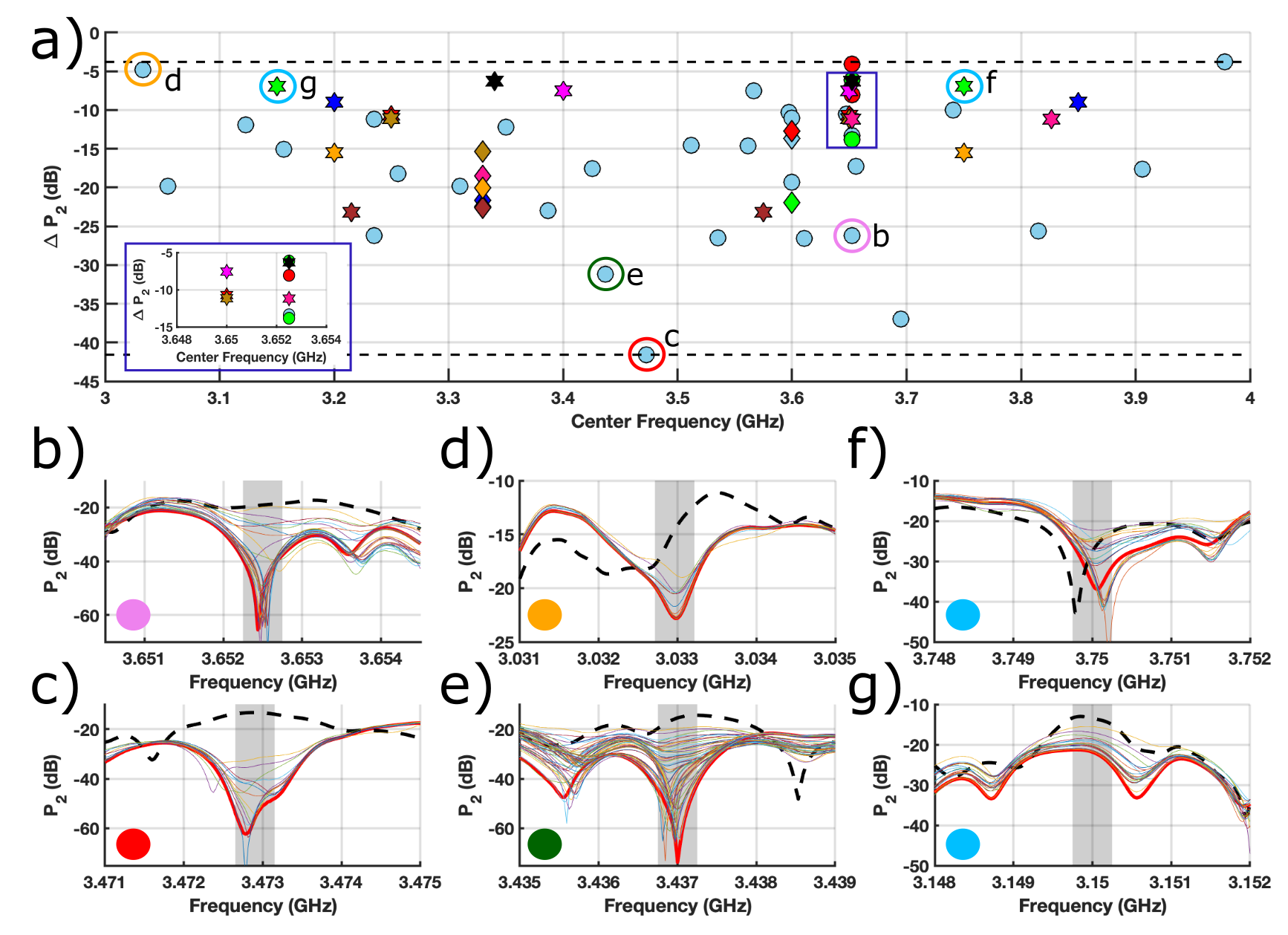}
\caption{\label{fig:power_optimization} \bf Results of minimizing power at port 2 with the iterative optimization algorithm. \bf a) \normalfont Plot of $\Delta P_2$ at various frequencies in the range of operation of the metasurface. Circles represent cases where the metric was evaluated over a single frequency band and are color coded by bandwidth (sky blue is 500 kHz, green is 5 MHz, and red is 10 MHz). Hexagrams represent dual frequency band metrics and are color coded by matching pairs. Diamonds represent driving both ports 1 and 3 collectively and are color coded by relative phase shift (0, 8, 15, 25, and 50 deg/GHz). Letters indicate points shown in detail in the following panels.  The dashed black lines indicate the smallest reduction (-4dB) and largest reduction (-41 dB). \bf b \normalfont through \bf g) \normalfont $P_2$ evolution from initial (all 0s) to final state. The dark shaded region represents the frequency band where $\Delta P_2$ was evaluated, the dashed black line shows the initial response, the solid bold red line shows the final response, and the remaining lines show a few of the incremental steps. \bf b \normalfont and \bf c) \normalfont Single band examples centered at 3.033 GHz and 3.6525 GHz, with 28 and 5 dB of suppression, respectively. \bf d \normalfont and \bf e) \normalfont Single band examples centered at 3.473 GHz and 3.437 GHz, with 41 and 31 dB of suppression, respectively. \bf f \normalfont and \bf g) \normalfont Dual band example centered at 3.75 GHz and 3.15 GHz, with 7 dB of suppression averaged over the 2 bands. }
\end{figure*}

Figure \ref{fig:power_optimization} shows the results obtained when minimizing the average power at the output port and compares the results of many different experiments and configurations.  All the cases are scored by the change in average power, $\Delta P_2$, between the initial inactive (all 0s) state and the final state. The optimization algorithm was performed with $\Delta P_2$ evaluated over a single frequency band as well as simultaneously over multiple separated frequency bands. As discussed previously, the widths of the nulls were observed to be $\sim$200 kHz. The initial bandwidth was selected to be 500 kHz in order to ensure that $\Delta P_2$ was evaluated over an entire null. In addition, the cavity configuration was switched between driving a single input port and driving two input ports simultaneously with varying relative phase shifts. The achieved suppression ranges from 4-40 dB with most cases providing $>$ 10 dB. The lower values of $\Delta P_2$ arise in the following cases: working near the edges of the metasurface operational window, evaluating $\Delta P_2$ over a large bandwidth, or evaluating $\Delta P_2$ over multiple separated bands.  This is not surprising as more bandwidth results in more features in the region where $\Delta P_2$ is evaluated, which then means more degrees of freedom are required to be manipulated for destructive interference. The metasurface provides some benefit outside of the 3-3.75 GHz design window; the reflection phase change of the pixels is limited near the edges of the operational bandwidth, so performance is expected to be reduced under those conditions. More details on specific cases are provided in the supplemental material.

Since $\Delta P_2$ is inherently a relative measurement, there is an implicit dependence on the initial state. Using the inactive (all 0s) state as the reference ensures the metasurface is always initialized with the same command even though the specific value is dependent on the selected frequency window. Starting with a condition where there was already a deep null would result in limited improvement; the average power in that case would already be quite low and there would not be a need for further reduction. Starting with a condition where there is a transmission peak however, would result in significant reduction. When using a single frequency band metric, we were able to drive deep nulls in each of the windows that were tested, as can be seen by the circles in Fig. \ref{fig:power_optimization}a and the power at port 2 in panels b) through e). Panels b) and d) show moderate cases where there is not a clear peak in the initial $P_2$ measurement, while panels c) and e) show cases with a clear peak in the initial $P_2$ measurement and demonstrate significant improvement. This highlights the dependence of $\Delta P_2$ on the initial state.

Deep transmission nulls were also observed when driving two input ports simultaneously with varying relative phase shifts, as shown by the diamonds in Fig. \ref{fig:power_optimization}a. This indicates our approach is self-adaptive and can compensate for multiple input signals as well as signals coming in from different directions. With dual frequency bands however, we were generally unable to drive deep nulls in both bands simultaneously, which can be seen by the hexagrams in Fig. \ref{fig:power_optimization}a and the power at port 2 over the frequency band in panels f) and g). This is because the metasurface frequency response in separated bands is correlated, as the metasurface induces wide bandwidth effects on the scattering properties of the enclosure. As shown in the supplemental material, different choices of metrics produce different out of band behavior. These results show that our approach provides 3 distinct advantages over previous works: 1) we are able to generate coldspots at arbitrary frequencies and are not limited to a single operating frequency; 2) we are able to generate coldspots simultaneously in multiple separated frequency bands as well as at single frequencies; and 3) we are able to generate coldspots when the injected signal comes from multiple directions with an arbitrary relative phase shift and are not limited to a single direction. 

\section{\label{sec:cpaState.png}Generating Coherent Perfect Absorption}
Coherent Perfect Absorption (CPA) is a situation in which all energy injected into a system is absorbed, no matter how small the losses are in the system. Creating CPA requires coherent excitation of all the ports in an eigenvector whose corresponding $S$-matrix eigenvalue is zero. Operationally, the first step in establishing CPA is to find an eigenvalue of the scattering matrix that is close to zero. For example, a 2 x 2 scattering matrix will have a pair of eigenvalues at each frequency. However, realizing CPA only requires driving 1 eigenvalue to zero, as the other eigenvalue corresponds to the anti-CPA state \cite{chenPerfectAbsorptionComplex2020}. For the following discussion and experimental results, we only consider the smallest eigenvalue of each pair.  
 
CPA has typically been investigated in simple, regular scattering scenarios and cavities but recently it has been demonstrated in more complex systems, specifically in the realm of wave chaos, and graphs \cite{kottosQuantumChaosGraphs1997, kottosChaoticScatteringGraphs2000, fyodorovDistributionZerosMatrix2017, liRandomMatrixTheory2017}. These works analytically demonstrate the use of RMT to explore CPA states with semiclassical tools without relying on the limit of weak coupling. CPA states have also been experimentally investigated in multiple scattering environments \cite{pichlerRandomAntilasingCoherent2019}, and in graphs that break time-reversal invariance \cite{chenPerfectAbsorptionComplex2020}. The use of enhanced spatio-temporal diversity from a metasurface for realization of CPA has not yet been explored.
 
Recent research however has investigated the use of metasurfaces for Perfect Absorption (PA) inside a cavity and demonstrated a secure communication system as an application \cite{imaniPerfectAbsorptionMetasurfaceProgrammable2020}. PA is a complementary idea to CPA for a single port system that relies only on the reflection coefficient, $S_{11}$ \cite{asadchyBroadbandReflectionlessMetasheets2015}. Coherent excitation of a single port with complete absorption has been demonstrated to enhance wireless power transfer \cite{krasnokCoherentPowerTransfer2018}. Our work extends this to coherent operation with the full scattering matrix for a 2-port system, and can be generalized to an arbitrary number of ports.
 
Realizing a true absolute zero of the $S$-matrix eigenvalues is generally difficult, because the eigenvalues are complex numbers. Thus two parameters must be varied independently to drive an eigenvalue to zero.  Further, a CPA state is highly dependent on the structure of the underlying scattering system. This is best understood in the framework of the Random Coupling Model (RCM) \cite{zhengStatisticsImpedanceScattering2006b, zhengStatisticsImpedanceScattering2006}. The eigenvalues accessible by means of the programmable metasurface tend to cluster around values determined by the coupling properties of the ports, which are characterized by the radiation $S$-matrix, $S_{\text{rad}}$. We define $S_{\text{rad}}$ as the $S$-matrix corresponding to the free-space radiation condition with the cavity walls taken out to infinity such that no waves come back to the ports \cite{hemmadyUniversalPropertiesTwoport2006}. $S_{\text{rad}}$ can be determined by a number of means \cite{yehWAVECHAOTICEXPERIMENTS2013}. Here we employ the ensemble average of the time gated measured $S$-parameters in the cavity \cite{addissieExtractionCouplingImpedance2019}, as described in the supplemental material. Deviations of the scattering matrix from $S_{\text{rad}}$ have a number of causes.  First there are deviations resulting from relatively direct ray paths between the ports \cite{yehUniversalNonuniversalProperties2010}.  These deviations are removed by averaging the $S$-matrix over a frequency window that is the reciprocal of the time of flight on the path. However, in finding the eigenvalues of the $S$-matrix in a narrow frequency range, these deviations are present. Second, there are deviations due the multitude of longer paths, and these are characterized statistically by RMT within the RCM.  These fluctuations in $S$ tend to be of the order of $1/(\pi \alpha)^{1/2}$ \cite{hemmadyUniversalImpedanceFluctuations2005, yehUniversalNonuniversalProperties2010,gradoniPredictingStatisticsWave2014} where the loss parameter $\alpha = f/(2\Delta fQ)=3$ in the present experimental case.  Finally, there are deviations dependent on the state of the metasurface.  These deviations are constrained to be less than or equal to either the direct path or the statistical long path deviations.

Thus, to find a CPA state it is necessary for the ports to be sufficiently matched so that the statistical fluctuations can shift the eigenvalues to zero. If the ports are poorly matched and losses within the cavity are sufficiently high, the eigenvalues will naturally fall near values determined by the properties of the ports with statistical fluctuations around those values dictated by the amount of cavity loss. As such, it is generally not possible to realize a CPA state at arbitrary frequencies when limited to a single DOF \cite{chenPerfectAbsorptionComplex2020}. The availability of additional DOF, such as those produced by the metasurface, allows greater control over the underlying scattering system and provides a greater likelihood of potential CPA states.

\begin{figure}
\includegraphics[width = 8.6 cm]{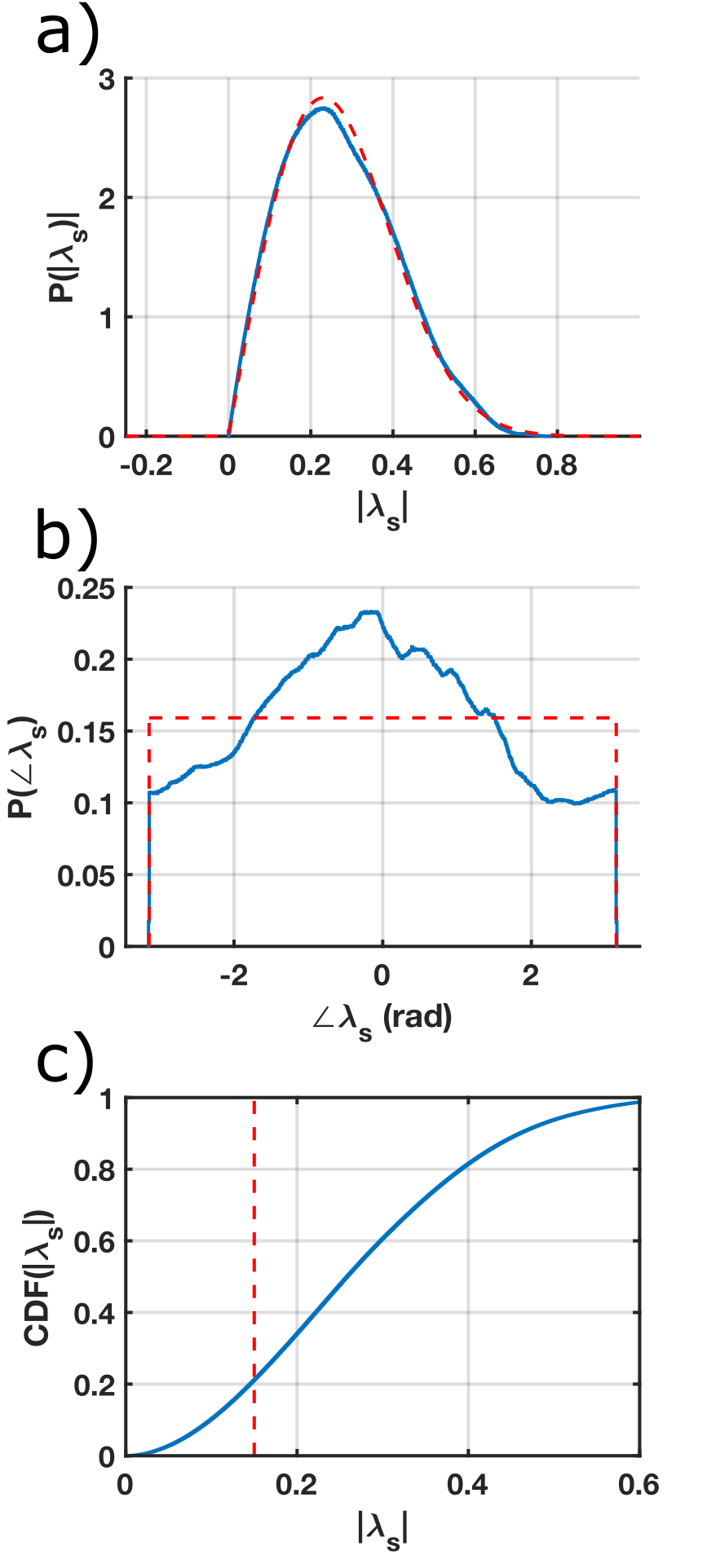}
\caption{\label{fig:eigenvalue_stats} \bf $S$-matrix statistics for a random distribution of 2000 metasurface commands. \normalfont Panels show statistics for the scattering matrix eigenvalues covering all commands and all frequencies from 3-4 GHz. \bf a) \normalfont PDF for the magnitude of scattering matrix eigenvalues, $|\lambda_s|$. The dashed red line shows the fit to a Rician distribution with $\sigma$ = 0.173 and $\nu$ = 0.177. \bf b) \normalfont PDF for the phase of scattering matrix eigenvalues, $\angle \lambda_s$. The dashed red line shows the distribution for a perfectly uniform phase. \bf c) \normalfont CDF for the magnitude of scattering matrix eigenvalues. The dashed red line shows the threshold of $|\lambda_s| <$ 0.15.}
\end{figure}

\begin{figure*}
\includegraphics[width = 17.2 cm]{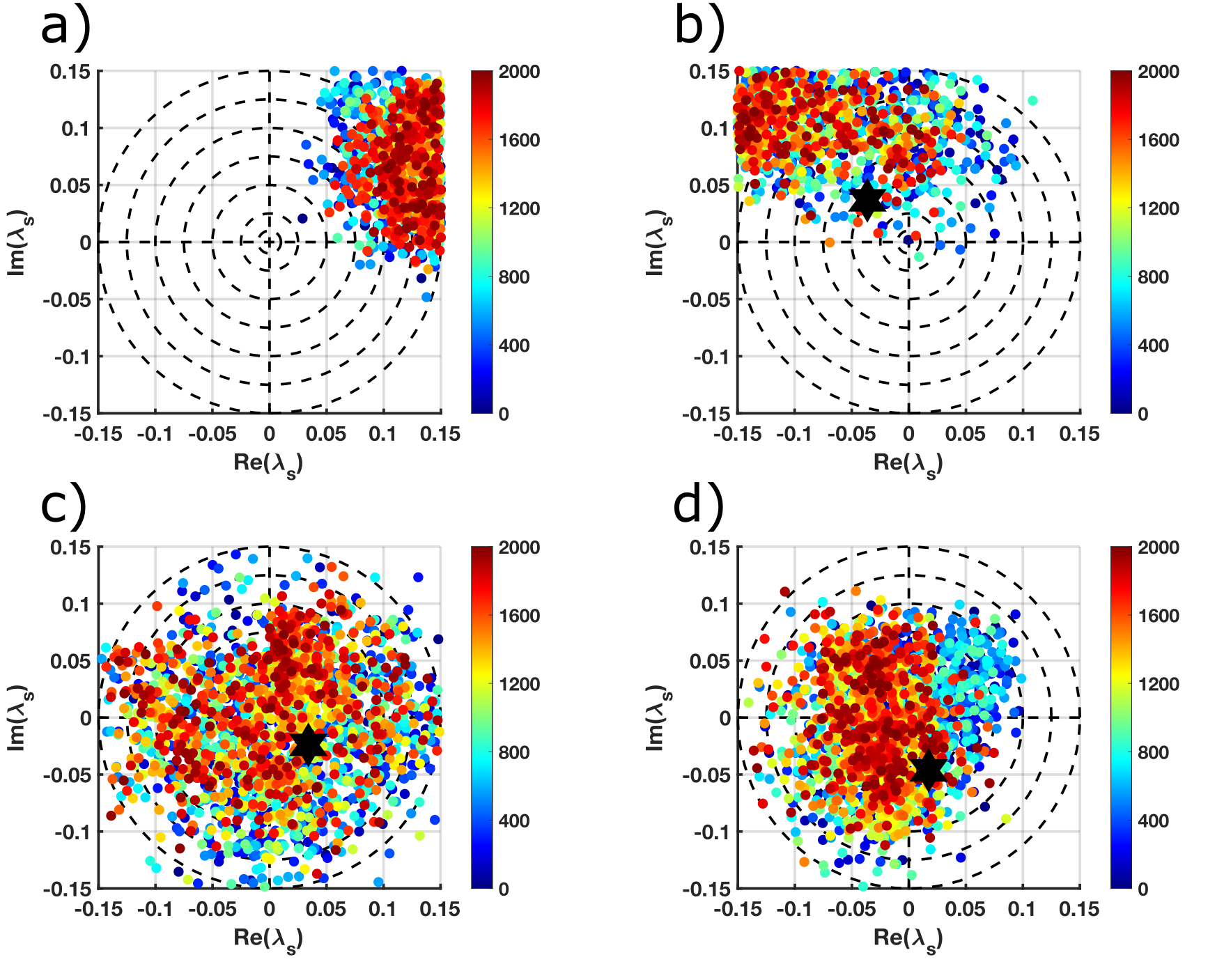}
\caption{\label{fig:eigenvalue_char} \bf Point clouds of selected $S$-matrix eigenvalues for a random distribution of 2000 metasurface commands. \normalfont Panels show the point clouds of the smaller eigenvalues of the 2000 scattering matrices at 4 selected frequencies. The colored circles are the $S$-matrix eigenvalues, and are color coded by the specific command, from 1 to 2000. The large black hexagrams indicate the position of the eigenvalue of $S_{\text{rad}}$. \bf a) \normalfont Candidate at $f$ = 3.0055 GHz, minimum $|\lambda_s|$ = $6\times10^{-2}$. \bf b) \normalfont Candidate at $f$ = 3.4021 GHz, minimum  $|\lambda_s|$ = $2\times10^{-3}$. \bf c) \normalfont Candidate at $f$ = 3.6564 GHz, minimum  $|\lambda_s|$ = $5\times 10^{-3}$. \bf d) \normalfont Candidate at $f$ = 3.9991 GHz, minimum  $|\lambda_s|$ = $5 \times 10^{-4}$.}
\end{figure*}

Characterization of the $S$-matrix eigenvalues from a distribution of 2000 command sets is presented in Figures \ref{fig:eigenvalue_stats} and \ref{fig:eigenvalue_char}. Figure \ref{fig:eigenvalue_stats} shows the probability distributions for all of the $S$-matrix eigenvalues over all frequencies and commands. Panel a) shows that the magnitude follows a Rician distribution as predicted by Ref. \cite{yehFirstprinciplesModelTimedependent2012}, which also tells us that the $\nu$ parameter of the Rician distribution is due to the presence of persistent short orbits \cite{hartEffectShortRay2009}. Panel b) shows that the phase of the $S$-matrix eigenvalues is not truly uniformly distributed. The deviation of the eigenphase from uniformity indicates that the random distribution is not statistically independent and again tells us there are persistent short orbits present in the system. These short orbits are not captured explicitly in $S_{\text{rad}}$, and will cause the eigenvalues of $S_{\text{rad}}$ to be offset from the center of the point cloud of $S$-matrix eigenvalues. Short orbits can be explicitly included analytically in the RCM \cite{hartEffectShortRay2009, yehExperimentalExaminationEffect2010} and do not prevent us from proceeding. Panel c) shows the CDF of the eigenvalue magnitudes and is useful in establishing thresholds for potential CPA candidates.

Figure \ref{fig:eigenvalue_char} shows point clouds of the $S$-matrix eigenvalues at selected frequencies and demonstrates that the eigenvalues can have very different behavior in how they approach the origin. The panels show the collection of eigenvalues of the 2000 $S$-matrices at selected frequencies along with the eigenvalues of $S_{\text{rad}}$ at that frequency. We can see that the eigenvalues of the distribution tend to cluster around the eigenvalues of $S_{\text{rad}}$; the offset from the center of the point cloud is due to the presence of short orbits, as discussed above. In panel a), the $S$-matrix eigenvalues are clustered in the upper right quadrant far from the origin and do not enter the inner rings. The $S_{\text{rad}}$ eigenvalue is in the upper right-hand quadrant outside of the plot area, at $0.1862+j0.2288$. In panel b), the $S$-matrix eigenvalues are clustered in the upper half, with some getting close to the origin. In panel c), the $S$-matrix eigenvalues show a fairly uniform density throughout the full $|\lambda_s| <$ 0.15 range. In panel d), the $S$-matrix eigenvalues show a high density clustered around the origin. The results in these panels are from a random distribution of commands rather than a targeted search. During optimization, we will take smaller dithering steps for finer control as we approach the origin, and expect to see slightly different behavior. 

The variance in eigenvalue magnitudes means we need to use a large threshold for identifying candidates because the overall global minimum $S$-matrix eigenvalue may not be identified as a candidate in every realization. In practice, we found that we were unable to realize CPA states when starting with a magnitude $|\lambda_s|\geq$ 0.2 but were generally able to realize CPA states when starting with a magnitude $|\lambda_s|\leq$ 0.15. Moving an eigenvalue far from the origin requires modifying the underlying scattering matrix more strongly than moving an eigenvalue that is already near the origin, so this behavior is expected.  Assessing the probability of finding a CPA state in a given frequency range a priori is difficult. The universal properties of a complex scattering system are not easily separated from the deterministic properties when working with $S$-parameters, as the statistics are dominated by $S_{\text{rad}}$ \cite{hemmadyUniversalStatisticsScattering2005}. This means the existence of a CPA state is highly dependent on the coupling properties of the ports and therefore the specific antennas chosen. An analytical approach is possible through the framework of the RCM and will be left to future work.

An open question is how small do the eigenvalues need to be to realize CPA? This is dependent on the specific application and scattering system, as that determines how accurately the eigenvalues can be measured and maintained. For our experimentation, we set $|\lambda_s| \leq 5\times10^{-3}$ as the upper bound and $|\lambda_s| \leq 1\times10^{-3}$ as the goal for realizing CPA.

\begin{figure*}
\includegraphics[width = 17.2 cm]{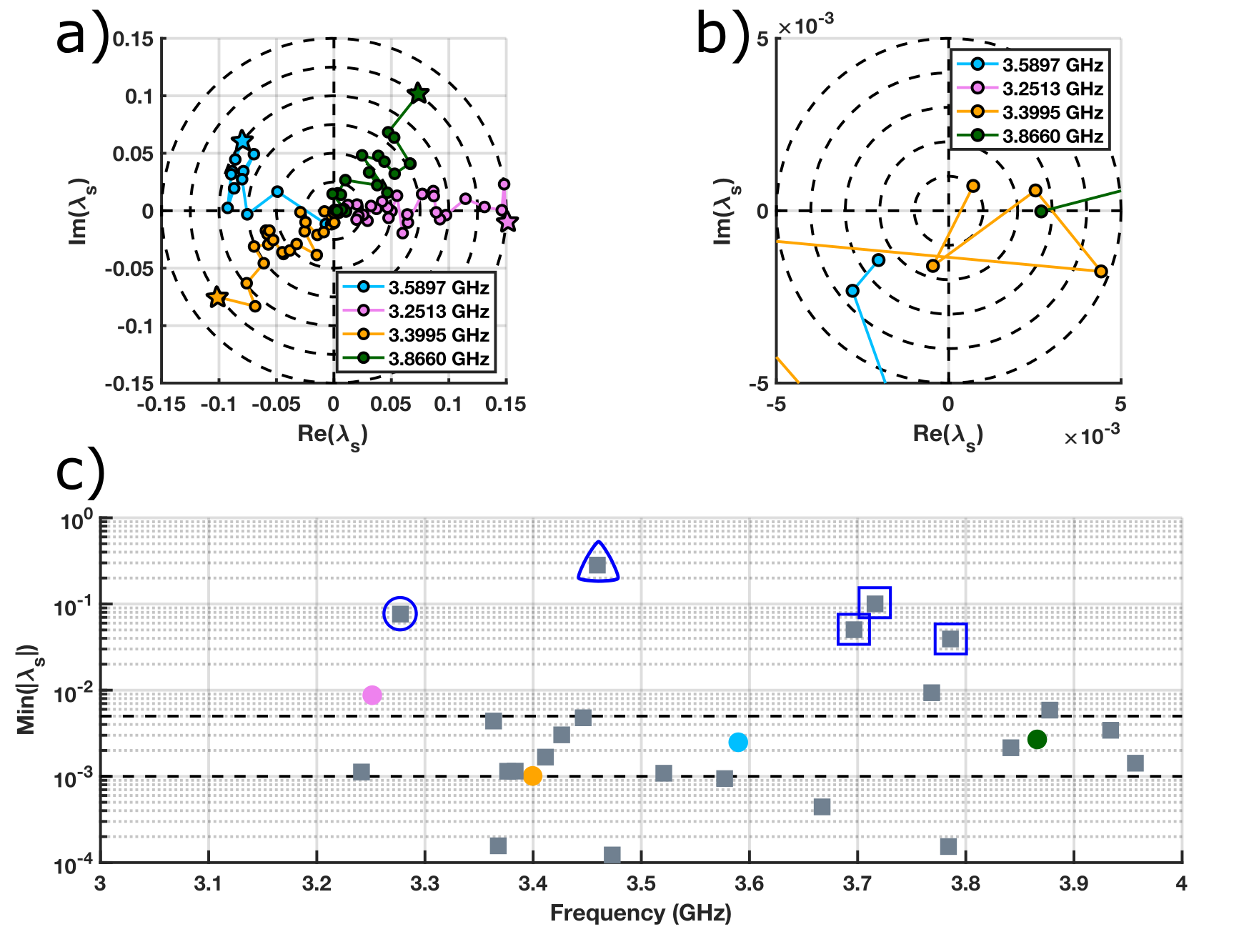}
\caption{\label{fig:cpa_state} \bf Experimental $S$-matrix eigenvalue trajectories for realization of Coherent Perfect Absorption (CPA) states. a) \normalfont and \bf b) \normalfont Directed trajectories for eigenvalues showing the random walk nature of the algorithm and demonstrating mobility of selected eigenvalue candidates. \bf a) \normalfont  Zoomed out view showing selection of initial $S$-matrix eigenvalue candidates and behavior away from the origin, the bulls-eye circles are spaced at radii incrementing by $2.5\times10^{-2}$. The starting eigenvalue magnitude in each case is identified by a star. \bf b) \normalfont Close up view showing behavior near the origin, the bulls-eye circles are spaced at radii incrementing by $1\times10^{-3}$. Of the four cases shown, only 3 were able to get inside the inner rings near the origin where $|\lambda_s| < 5\times10^{-3}$. \bf c) \normalfont Minimum achieved eigenvalue magnitude for each performed experiment. The circles indicate data that is shown in the upper plots and are color coded to match. The gray squares indicate an experiment that was performed but whose detailed trajectory is not shown in the upper plots. The dashed black lines indicate the cross over points of $5\times10^{-3}$ and $1\times10^{-3}$. The enclosed squares indicate cases where the initial eigenvalue magnitude $|\lambda_s| >$ 0.15. $|\lambda_s| $ = 0.5 for the triangle, $|\lambda_s| $ = 0.2 for the squares, and $|\lambda_s| $ = 0.175 for the circle.}
\end{figure*}

We adopt the same basic algorithm used for power minimization but initialize it differently. We apply a random set of commands to the metasurface and then select a candidate eigenvalue with a specified magnitude.  Figure \ref{fig:cpa_state} presents the results of 27 separate CPA eigenvalue optimization experiments. Fig. \ref{fig:cpa_state}a shows the behavior of 4 selected cases away from the origin for $|\lambda_s| < 0.15$, and Fig. \ref{fig:cpa_state}b shows the behavior at the CPA condition for $|\lambda_s| < 5\times10^{-3}$. Only 3 of the 4 selected cases reach the CPA threshold. Fig. \ref{fig:cpa_state}c presents the collection of all 27 experiments, with the four shown in detail in panels a) and b) color coded. Each case was initialized with an eigenvalue magnitude chosen in the range $0.075 \leq |\lambda_s| \leq 0.5$. The case that started with $|\lambda_s| = 0.5$ is enclosed by a triangle, the cases that started with $|\lambda_s| = 0.2$ are enclosed by squares and the case that started with $|\lambda_s| = 0.175$ is enclosed by a circle. All the rest started with $|\lambda_s| \leq 0.15$. Three cases initialized with $|\lambda_s| = 0.15$ did not quite make the CPA threshold, $|\lambda_s| \leq 5\times10^{-3}$. Two cases were within a factor of 2, $|\lambda_s| \leq 9\times10^{-3}$, while the third was within $\sim$20\%, $|\lambda_s| = 6\times10^{-3}$ . 

Utilizing the iterative optimization algorithm to change the metasurface, we are able to drive eigenvalues towards the origin in all cases, but the algorithm stalls at different points. The closer we get to the origin, the more difficult it becomes to reduce the eigenvalue further. As with the coldspot optimization, the stochastic nature of the algorithm plays a role in where convergence is reached. The overall performance could be improved by increasing the convergence criteria or making the algorithm adaptive so that it tracks multiple candidates and switches to another candidate when the optimization stalls.

As a final step, we want to verify that the CPA state has been achieved. Because the CPA state is found by minimizing the eigenvalues of the scattering matrix, verification requires that we apply the corresponding $S$-matrix eigenvector. This can be done using a network analyzer with 2 independent sources and an external phase shifter \cite{chenPerfectAbsorptionComplex2020}. After directing a particular eigenvalue towards the origin, the network analyzer was configured for independent source operation and the amplitude and phase were adjusted to generate the eigenvector, as described in the supplemental material. The presence of a CPA state is verified by looking at the ratio of all the power emerging from the cavity to all the power injected into the cavity, $P_{\text{out}}/P_{\text{in}}$. Sensitivity to changes in the eigenvector can be determined by making small deviations in the relative phase shift or amplitude between the two sources. Sensitivity to the eigenvalue can be determined by small changes in frequency. 

\begin{figure*}
\includegraphics[width = 17.2 cm]{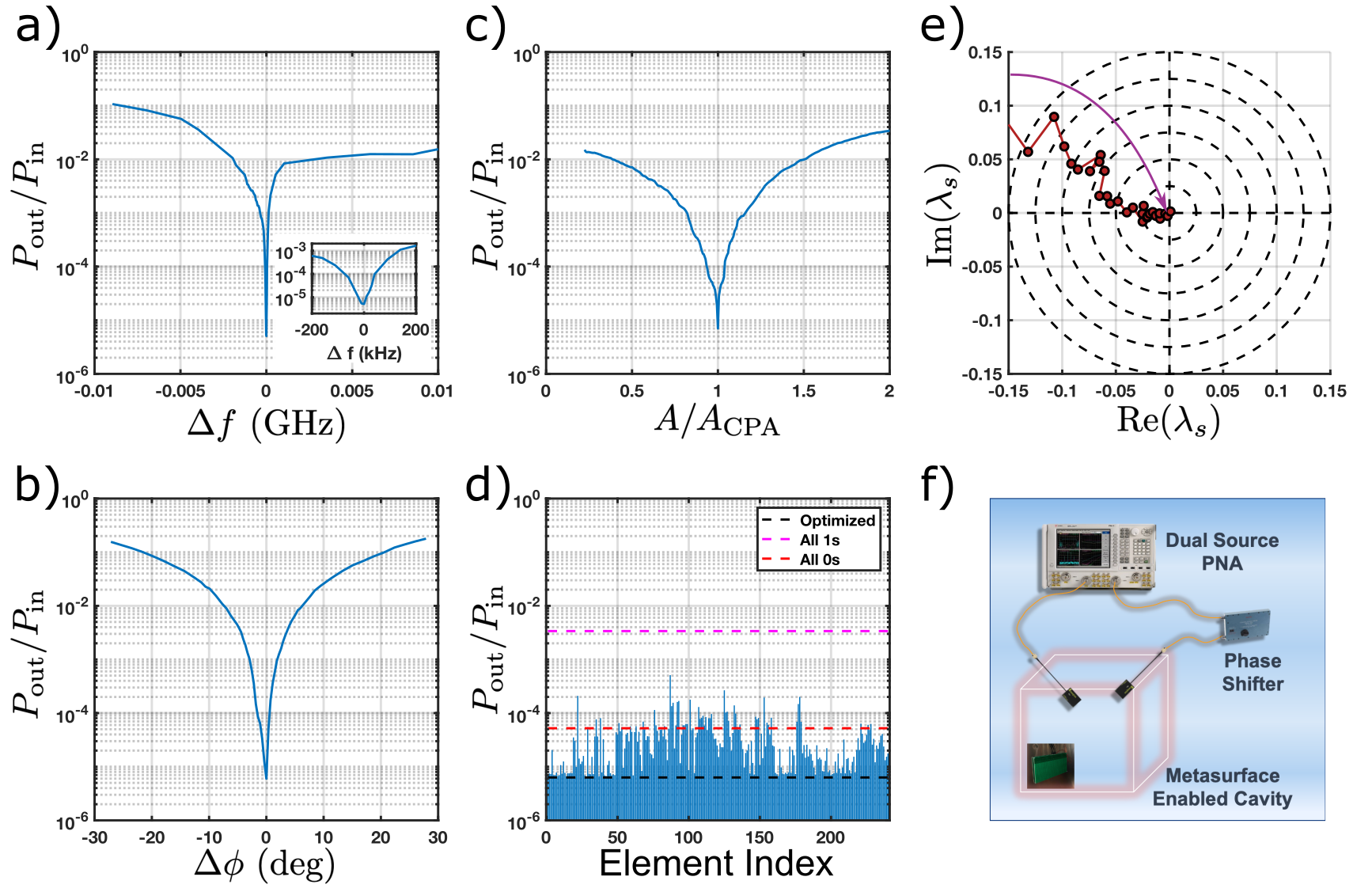}
\caption{\label{fig:cpa_sweeps} \bf Coherent Perfect Absorption (CPA) state verification at 3.6697 GHz. a) \normalfont Frequency sweep showing the power ratio, $P_{\text{out}}/P_{\text{in}},$ over a $\pm$ 10 MHz window, with the inset providing a closeup of the null in a $\pm$ 200 kHz window. \bf b) \normalfont $P_{\text{out}}/P_{\text{in}}$ vs. phase difference, $\Delta \phi$, showing the power ratio over a $\pm$ 30$^{\circ}$ window. \bf c) \normalfont $P_{\text{out}}/P_{\text{in}}$ vs. relative amplitude showing the power ratio when driving port 1 with an amplitude $\sim$0-2 times the CPA amplitude ($\text{A}_{\text{CPA}}$). \bf d) \normalfont Metasurface command sweep showing the power ratio when toggling individual elements relative to the optimized set. Each bar indicates the power ratio when that particular element was flipped between a 1 or a 0. The black dashed line shows the power ratio for the optimized state, the red dashed line shows the power ratio for the all 0s state, and the magenta dashed line shows the power ratio for the all 1s state. \bf e) \normalfont Eigenvalue magnitude trajectory during optimization of the CPA state prior to performing the verification sweeps. Minimum achieved $|\lambda_s| = 4\times10^{-4}$. \bf f) \normalfont Diagram showing experimental setup for applying CPA eigenvector excitation and verification sweeps.}
\end{figure*}

A set of parameter sweeps that verify a CPA state was realized are presented in Figure \ref{fig:cpa_sweeps}. Fig. \ref{fig:cpa_sweeps}e shows the $S$-matrix eigenvalue magnitude trajectory during optimization prior to performing the verification sweeps. The overall experimental setup is shown in Fig. \ref{fig:cpa_sweeps}f, which shows that a 2-source network analyzer was configured with independent source operation and connected to the cavity with an external phase shifter on port 1. This allows us to produce the appropriate eigenvector by controlling the relative amplitude with the network analyzer and the relative phase with the phase shifter. The metric for the sweeps is the power ratio, $P_{\text{out}}/P_{\text{in}}$, of all the power emerging from the cavity to all the power injected into the cavity. At the CPA condition, all the energy should be absorbed. However, due to instrumentation limitations with the system noise floor, the smallest measurable power ratio is $\sim$10$^{-6}$. Before performing the sweeps, the eigenvector was tweaked to provide the closest CPA state realization and then the parameters were varied to determine the sensitivity of the power ratio.

Fig. \ref{fig:cpa_sweeps}a shows the results of the frequency sweep performed in a $\pm$ 10 MHz window around 3.6697 GHz, with the inset showing a closeup in a $\pm$ 200 kHz window. The width of the deep null is $\sim$200 kHz, which matches the null widths found during cold spot generation. Fig. \ref{fig:cpa_sweeps}b shows the results of the phase sweep, which was performed by adjusting the external phase shifter. Here, $\Delta \phi$ represents the phase shift at port 1 away from the CPA eigenvector phase. Fig. \ref{fig:cpa_sweeps}c shows the results of the relative amplitude sweep. This was performed by sweeping the power injected into port 1 from -10 dBm to +10 dBm. The $x$-axis is then scaled to show the relative change in injected amplitude from the initial CPA state.  In each of these cases, the minimium power ratio is $\sim6\times10^{-6}$ and shows a steep cusp-like increase with the various parameters. Fig. \ref{fig:cpa_sweeps}d shows the results of the metasurface command sweep. In this case, the 240 individual metasurface elements were toggled to determine the impact of a single element on the CPA state. Several elements had negligible impact on the power ratio in comparison with the optimized value as seen in the dashed black line, but no toggles were found with clearly better performance. The elements in the center of the metasurface have a stronger impact than those at the edges of the metasurface, but the largest change from the CPA condition was observed by setting all the elements to 1s as shown in the dashed magenta line.

\section{\label{sec:discussion}Conclusions}
We have demonstrated the ability of a programmable metasurface to generate microwave coldspots in a chaotic cavity at arbitrary frequencies and showed this capability exists even when applied over multiple frequency bands simultaneously. The coldspots can be generated for different bandwidths and mulitple input port configurations that induce additional angular and spatial diversity. We have also utilized the programmable metasurface to control the eigenvalues of the scattering matrix and direct them towards the origin to realize a CPA state for the cavity. Finally, we verified the existence of a CPA state and demonstrated the sensitivity to parameter sweeps in frequency, phase, amplitude, and metasurface configuration. All of this is accomplished with a metasurface that covers only 1.5\% of the interior surface area of the cavity and a unique and effective stochastic algorithm to find desired outcomes despite the enormous space of possible metasurface commands.

Future research directions include quantitatively analyzing CPA in the framework of the Random Coupling Model \cite{zhengStatisticsImpedanceScattering2006, zhengStatisticsImpedanceScattering2006b, hemmadyUniversalImpedanceFluctuations2005, gradoniPredictingStatisticsWave2014}, and using deep learning to facilitate generating optimal metasurface commands to minimize power and/or realize CPA states.

\section{\label{sec:acknowledgements}Acknowledgements}
We thank Joseph Miragliotta, David Shrekenhamer, and Robert Schmidt for providing the metasurface and supporting detailed discussions on operational use. We also thank Tsampikos Kottos for insights into CPA, and Lei Chen for instructions on performing the various CPA verification sweeps. Funding for this work was provided through AFOSR COE Grant FA9550-15-1-0171 and ONR Grant N000141912481. 

\section{\label{sec:contributions}Author Contributions}
B. Frazier performed the experiments and processed the data under the supervision of T. M. Antonsen and S. M. Anlage. B. Frazier prepared the manuscript with input from all co-authors.

\section{\label{sec:compete}Competing Interests}
The authors declare no competing interests.

\section{\label{sec:materials}Materials and Correspondence}
Correspondence and requests for materials should be addressed to B. Frazier (email: benjamin.frazier@jhuapl.edu).

\section{\label{sec:data}Data Availability}
The data and analysis scripts that support results presented within this paper and other findings of this study are available from the corresponding author upon reasonable request.


%

\end{document}


\title{Supplemental Material for Wavefront Shaping with a Tunable Metasurface: Creating Coldspots and Coherent Perfect Absorption at Arbitrary Frequencies}

\author {Benjamin W. Frazier}
\email{Benjamin.Frazier@jhuapl.edu}
\affiliation{Institute for Research in Electronics and Applied Physics, University of Maryland, College Park, MD 20742, USA}
\affiliation{Department of Electrical and Computer Engineering, University of Maryland, College Park, MD 20742, USA}
\affiliation{Johns Hopkins University Applied Physics Laboratory, Laurel, MD 20723, USA}

\author{Thomas M. Antonsen, Jr.}
\affiliation{Institute for Research in Electronics and Applied Physics, University of Maryland, College Park, MD 20742, USA}
\affiliation{Department of Electrical and Computer Engineering, University of Maryland, College Park, MD 20742, USA}
\affiliation{Department of Physics, University of Maryland, College Park, MD 20742, USA}

\author{Steven M. Anlage}
\affiliation{Department of Electrical and Computer Engineering, University of Maryland, College Park, MD 20742, USA}
\affiliation{Quantum Materials Center, University of Maryland, College Park, MD 20742, USA}
\affiliation{Department of Physics, University of Maryland, College Park, MD 20742, USA}

\author{Edward Ott}
\affiliation{Institute for Research in Electronics and Applied Physics, University of Maryland, College Park, MD 20742, USA}
\affiliation{Department of Electrical and Computer Engineering, University of Maryland, College Park, MD 20742, USA}
\affiliation{Department of Physics, University of Maryland, College Park, MD 20742, USA}
\date{\today}
\date{\today}

\maketitle

\section{Experimental Setup}
\begin{figure*}
\includegraphics[width=17.2cm]{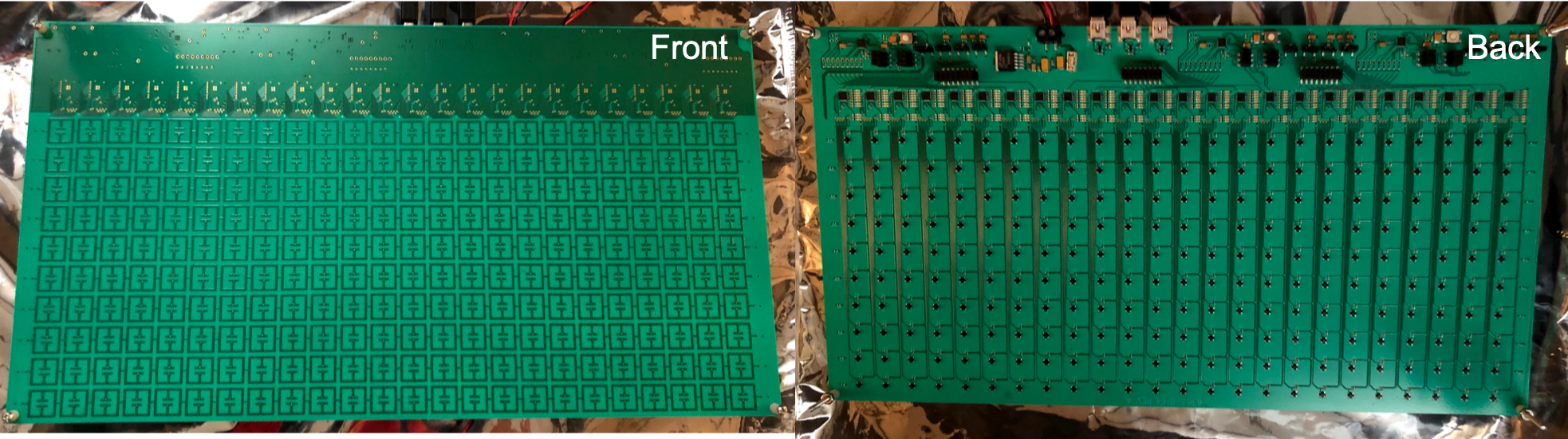}
\caption{\label{fig:reflectArrayDevice} \bf Metasurface device showing both the front and back of the board.}
\end{figure*}

The metasurface used for the experimentation is a reflectarray fabricated by the Johns Hopkins University Applied Physics Laboratory (JHU/APL) and is shown in Figure \ref{fig:reflectArrayDevice}. The individual LC resonator unit cells can be seen on the front side and the GaAs FET amplifiers can be seen on the back side. It is designed to operate in the S-band from 3-3.75 GHz and provides 240 binary unit cells arranged in a rectangular grid of 10 rows and 24 columns. The unit cells are electric LC resonators driven by GaAs FET amplifiers with characteristic length $< \lambda/6$ \cite{schmidSbandGaAsFET2020}. 

\begin{figure}
\includegraphics[width = 8.6cm]{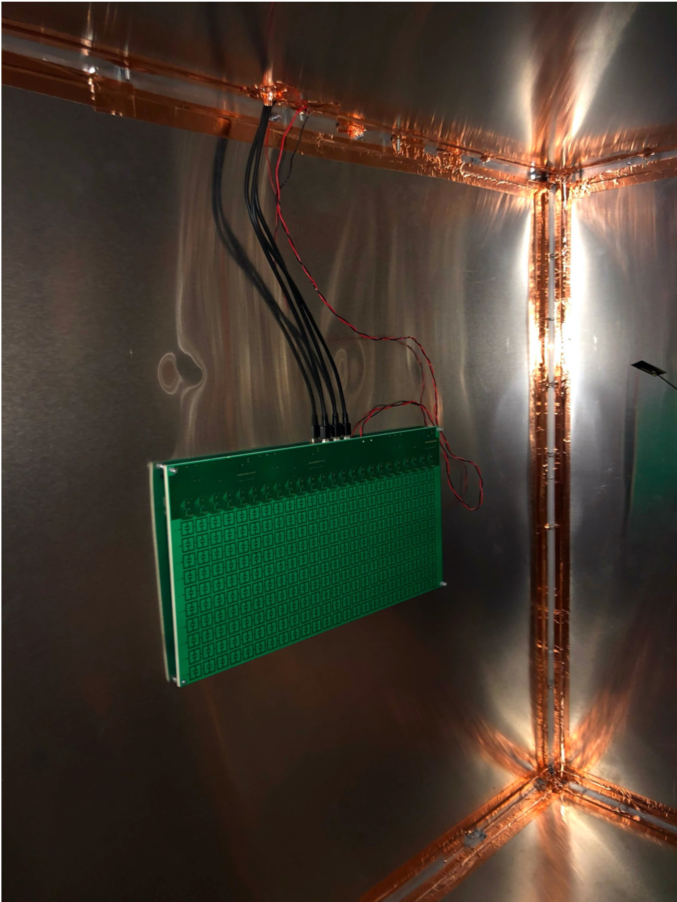}
\caption{\label{fig:reflectArrayInstalled} \bf Metasurface installed inside the cavity.}
\end{figure}

A test cavity was constructed using aluminum angle brackets for the frame and 0.019 inch thick aluminum sheets for the sides. Each side length is 3ft (0.9144 m), so the total cavity volume is 0.76 m$^3$ and the total surface area is 5.02  m$^2$. This geometry ensures the cavity is overmoded with at least 9 wavelengths per side, but unfortunately means that the active area of the metasurface only covers a small portion ($\sim 1.5 \%$) of the total surface area. All interior joints were lined with copper tape to minimize losses and hemispherical scatterers were installed in the corners of the cavity to force an irregular shape, after which the effective volume was reduced to 0.74 m$^3$. A higher loss version of the cavity was realized by distributing absorbing material in the bottom of the cavity. The metasurface was installed on a wall of the cavity as shown in Figure \ref{fig:reflectArrayInstalled}, with a 1/4 inch gap between the metasurface and the wall. This figure also shows the power and 3 USB cables running out through the top of the cavity.

\begin{figure}
\includegraphics[width=8.6cm]{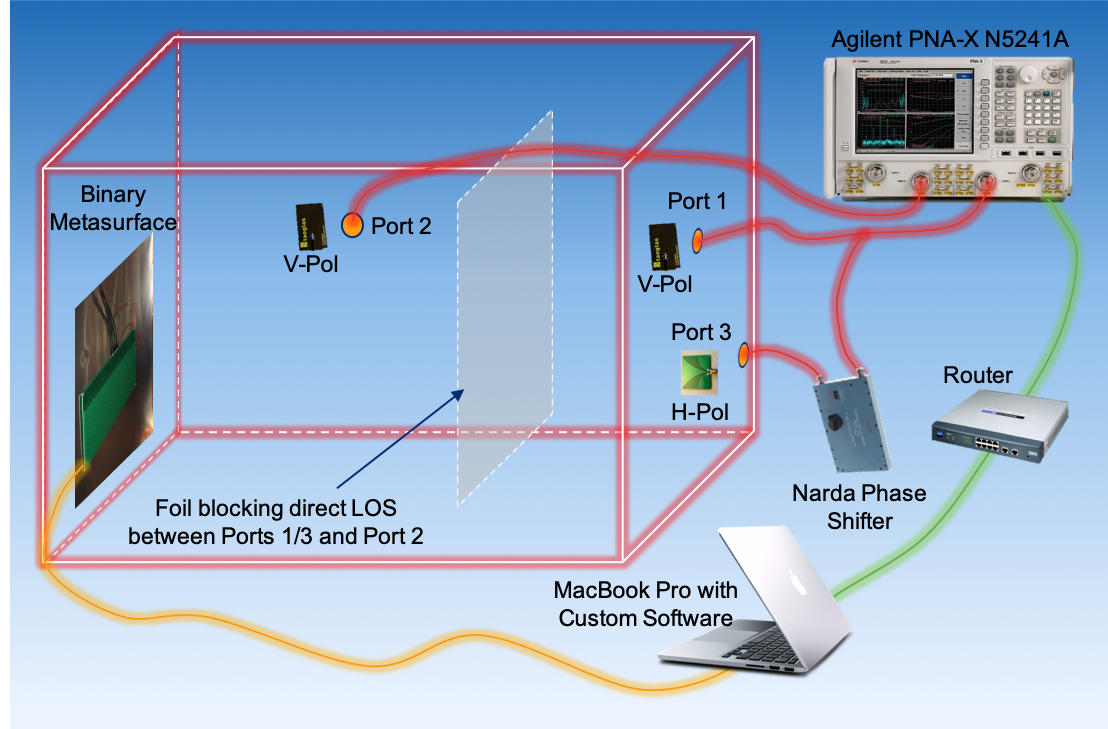}
\caption{\label{fig:cavityConfiguration} \bf Experimental setup.}
\end{figure}

An overview of the experimental setup is shown in Figure \ref{fig:cavityConfiguration} and shows the cavity configuration, the locations of the 3 ports relative to the metasurface, and the overall connectivity. The cavity has 3 ports with port 2 used for scoring and ports 1 and 3 used for signal injection. The input ports can be driven either individually or collectively with a relative phase shift. When they are driven collectively, the underlying scattering matrix is 3 x 3; this extra dimensionality is hidden when using a 2-port network analyzer. To ensure that wave interaction with the metasurface dominates the results, a sheet of foil is used to block the direct line of sight path from the input ports to port 2. Ultra Wideband (UWB) antennas designed for operation from 3-10 GHz are connected to each port. The antennas connected to ports 1 and 3 are Taoglas FXUWB10 antennas and are mounted so they radiate outward into the cavity from the walls in a vertically polarized configuration. The antenna connected to port 3 is a PCB module mounted orthogonally so that it is horizontally polarized.

Finally, the metasurface is controlled through 3 USB interfaces using custom C++ software with a Python wrapper on a MacBook Pro laptop, which also controls the Agilent network analyzer through an ethernet connection. In order to prevent corruption from noise, multiple measurements are averaged.

\section{Cavity Losses}
The cavity time constant, $\tau$, is an intrinsic aspect of the cavity that is dependent on losses but independent of the specifics of the underlying scattering system. This means $\tau$ is not dependent on the positioning or characteristics of the ports or antennas used to obtain it \cite{hollowayReverberationChamberTechniques2012}. $\tau$ is an important characteristic of the cavity as it is related to the quality factor, $Q$, through $Q = \omega \tau$. One way to estimate $\tau$ is through Power Delay Profile (PDP), which is the ensemble average of the magnitude squared of the Inverse Fourier Transform (IFT) of the transmission coefficient, $\text{PDP} = \left<\left|\text{IFT}\left\{S_{21}  \right\}   \right|^2  \right>$ \cite{hollowayThreeantennaTechniqueDetermining2012,hollowayEarlyTimeBehavior2012}. Since the power in the cavity decays exponentially, we can perform a linear fit on the logarithmic magnitude and estimate $\tau$ as $4.34/\nu$ where $\nu$ is the slope of the PDP in dB/s \cite{addissieExtractionCouplingImpedance2019}.

\begin{figure}
\includegraphics[width = 8.6cm]{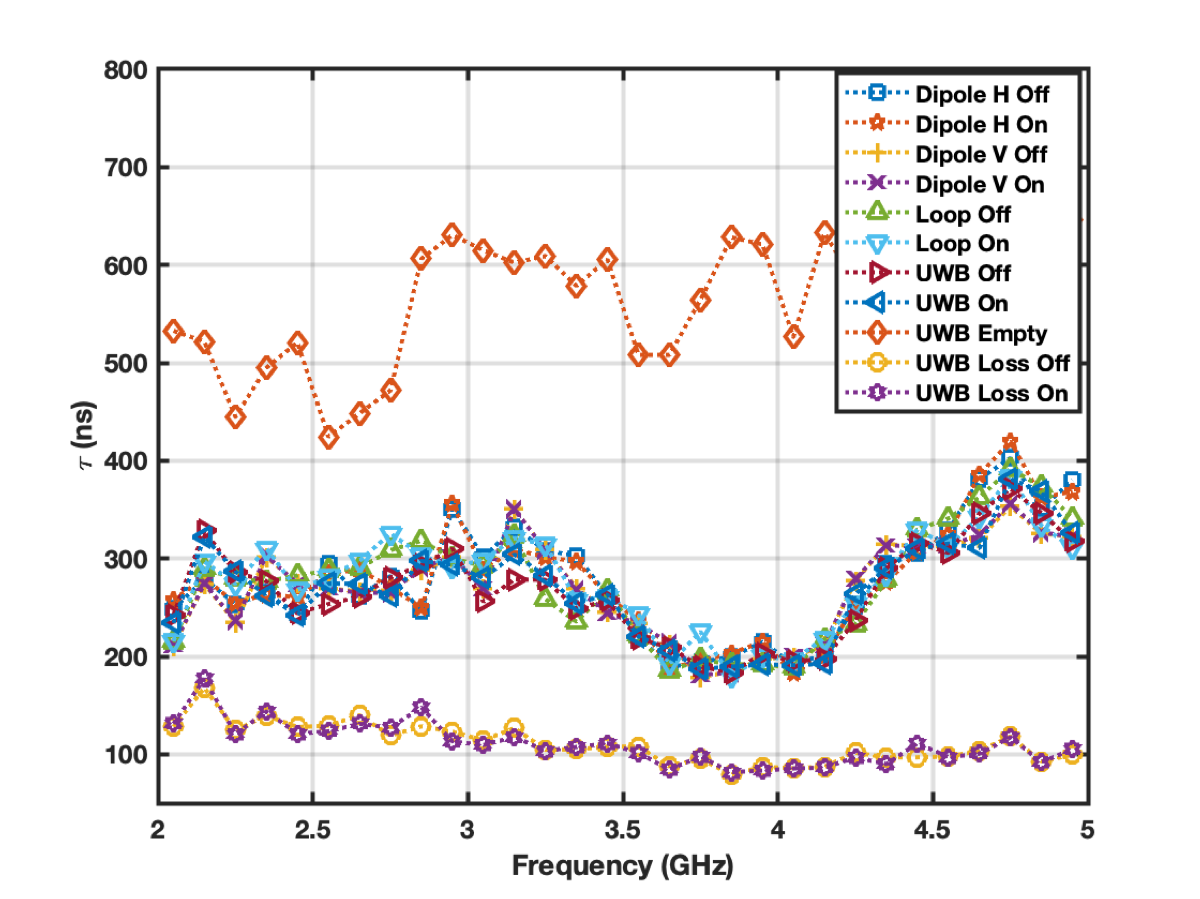}
\caption{\label{fig:cavity_tau} \bf Estimated cavity time constant for various configurations. \normalfont The three primary groupings are the empty cavity without the metasurface (top), the low loss configuration with the metasurface (middle), and the high loss configuration with the metasurface (bottom).}
\end{figure}

Figure \ref{fig:cavity_tau} shows the time constant estimated for the cavity under various configurations. There were 3 antenna pairs used in the PDP measurement: dipoles with both horizontal and vertical polarization, loops, and Ultra Wide Band (UWB). Measurements were taken with the cavity empty before installing the metasurface and in low and high loss cases with the metasurface installed. Each data point in Figure \ref{fig:cavity_tau} came from a 100 MHz window centered at the corresponding frequency. Installing the metasurface in the cavity had a significant impact on the time constant, reducing it by a factor of 2. Adding absorbing material for the high loss configuration reduced the time constant by another factor of 2. Powering on and off the metasurface however, had little impact on the time constant in either configuration. The PDP was measured between ports 1 and 2 with 3 different antenna pairs: dipole (in both horizontally and vertically polarized orientations), loop and Ultra Wide Band (UWB). The On and Off curves indicate whether the metasurface was powered on or off.

\section{Extracting Radiation $S$-Parameters}
In many cases, it is not feasible to directly measure $S_{\text{rad}}$, but it can extracted from time gating the $S_{\text{cav}}$ measurements \cite{addissieExtractionCouplingImpedance2019}. To ensure that the important features of $S_{\text{rad}}$ are maintained, the optimal gate width is determined by examining the $S$-parameters in the time domain and selecting the point in time where the individual ray trajectories start to diverge from the average.

The conventional time-gating approach is to convert the signal into the time domain with an Inverse Fast Fourier Transform (IFFT), multiply the signal with a rectangular gate, and then convert the signal back to the frequency domain with a Fast Fourier Transform (FFT). This approach compounds truncation effects through the IFFT/FFT sequence and induces band edge roll off effects due to the fact that we are using a finite, single-sided spectrum \cite{dunsmoreGatingEffectsTime2008}. An alternative approach is to perform gating in the frequency domain through a convolution operation and use the concept of renormalization to remove band edge artifacts. In order to optimize frequency domain gating, the gate needs to be centered at the time of the response being gated \cite{dunsmoreGatingEffectsTime2008}. Because we are interested in gating the initial response, we will center the time window at 0 seconds; accordingly, the overall width of the gate will be double the desired end time to include both positive and negative time extents.

The gate can be designed in 2 segments: a rectangular gate in time transformed to the frequency domain, and a window to reduce side lobes and ringing artifacts due to the sharp transitions of the rectangular gate. The generalized gate function in the frequency domain, $G(f)$, for a rectangular time domain gate defined between times $t_1$ and $t_2$ is given by a $\text{sinc}$ function, where we define $\text{sinc}(x) = \sin(\pi x)/ (\pi x)$.
\begin{equation}
G(f) = (t_2 - t_1)\text{sinc}\left[f\right(t_2 - t_1 \left) \right]\exp\left[-j2\pi f(t_2 + t_1) \right]
\end{equation}

For a gate centered at $t=0$ with end time $t_g$ ($t_1 = -t_g, t_2 = t_g$), this expression is simplified.
\begin{equation}
\label{eq_gate}
G(f) = 2t_g\text{sinc}\left[2 ft_g\right]
\end{equation}

The next step is to design a window function in the frequency domain to suppress side lobes. A common window is the Kaiser-Bessel window, which is defined by a shape parameter, $\beta$ and the window length, $M$ \cite{oppenheimDiscretetimeSignalProcessing2010}. For the analysis described here, the Agilent PNA provided 32,001 points of data and the window was defined with $\beta = 6.5$ and $M = 23,897$. We can then apply the gate by convolving the product of $G(f)$ and $W(f)$ with the measured $S$-parameter for a given cavity realization. Re-normalization is done by dividing the gated $S$-parameter by the convolution of the product of $G(f)$ and $W(f)$ with a constant unit frequency response, which removes band edge roll off effects \cite{dunsmoreGatingEffectsTime2008}. $S_{\text{rad}}$ is then found by taking the average of the results over the ensemble of cavity configurations as shown in Equation \ref{eq_apply_gate}.

\begin{equation}
\label{eq_apply_gate}
S^{\text{rad}}(f) = \left<\frac{\left[G(f) W(f)\right)]* S(f)}{\left[G(f) W(f)\right)]* 1(f)}\right>
\end{equation}

\section{Diverse Cavity Realizations}
Attempts were made to decompose the input commands into a deterministic set of orthogonal basis functions. This included driving single elements, columns of elements and even Hadamard matrices. Unfortunately, these all produced a narrow range of responses that did not cover the full extent of possibilities. A Hadamard matrix provides an orthonormal basis and decomposes sequences similarly to spatial frequencies; it is less computationally intensive than 2-D Fourier transforms and has many applications in multi-input multi-output communication theory and synthetic aperture imaging \cite{seberryApplicationsHadamardMatrices2005, trotsMutuallyOrthogonalGolay2015}.

\begin{figure*}
\includegraphics[width = 17.2cm]{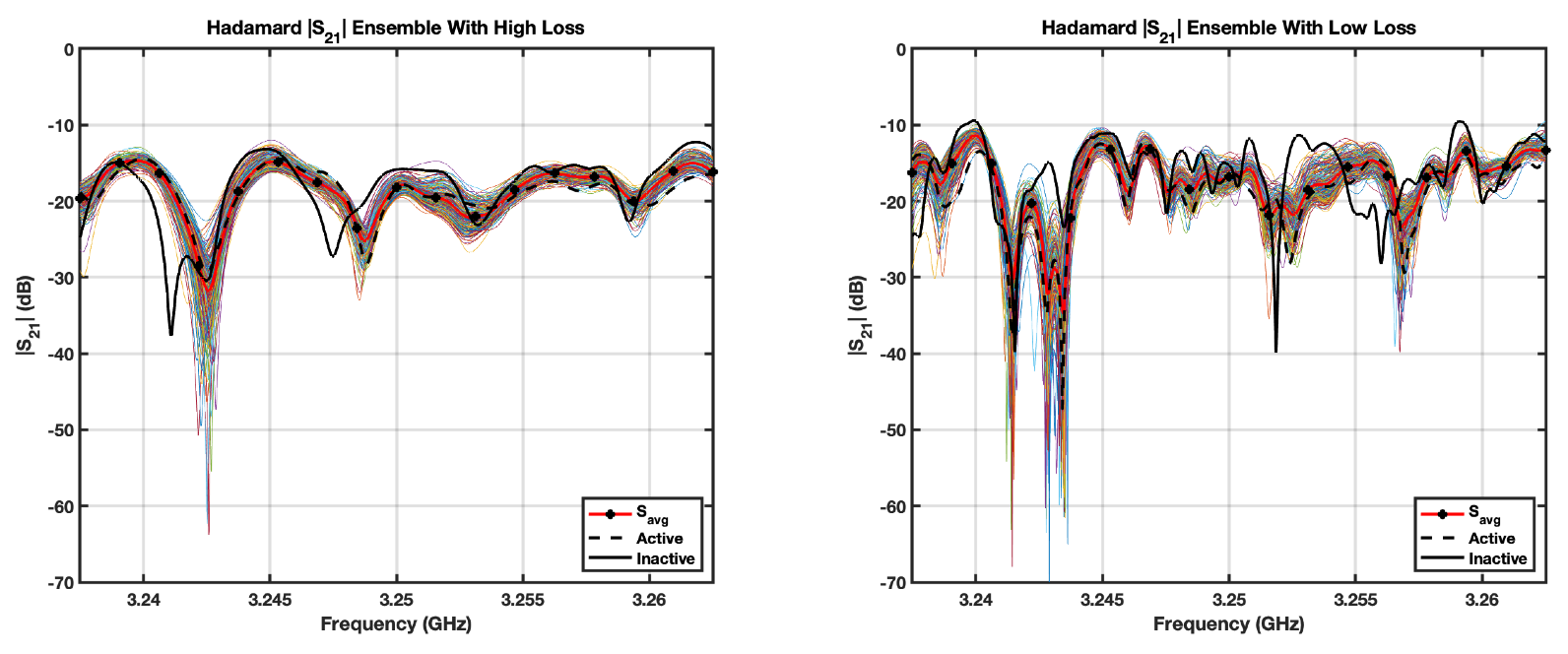}
\caption{\label{fig:hadamard_results} \bf Hadamard sequence $|S_{21}|$ ensembles for both high loss and low loss cases. \normalfont The high loss case is shown on the left hand side and the low loss case is shown on the right hand side.}
\end{figure*}

Figure \ref{fig:hadamard_results} presents the resulting ensembles from driving the metasurface with Hadamard basis functions and shows that the responses are narrow and do not cover the full extent. While the spatial frequency content and grouping of elements in the command sets changed, the number of active elements did not. An unbiased random coin toss approach likewise yielded a narrow distribution as roughly 1/2 the elements were active in each command set. The ensembles do not span the range covered by the inactive (all 0s) and active (all 1s) cases and do not cover the full extent. More variation is present in the low loss case because the ray trajectories survive longer and have more opportunities to interact with one another

The overall best approach to generating diverse ensembles was to use doubly random methods, in which compound probability distributions are used. This implies the statistics follow a Cox process, which is a generalization of a Poisson process with the underlying intensity or local mean itself a random process \cite{cressieStatisticsSpatiotemporalData2011}. A random biased coin toss, where the bias itself was selected from a random draw for each command set worked well but only excited high spatial frequencies on the metasurface. To include lower spatial frequencies, we added a doubly random inverse power spectrum approach, where the power spectrum is just a power law on spatial frequencies with the exponent a random draw. 

\begin{equation}
    C = \text{Re}\left(\text{IFFT}\left\{\left[\mathcal{N}(0,1) + j\mathcal{N}(0,1) \right]     \sqrt{\mathbf{f_r}^{\gamma}}\right\}\right)
\end{equation}

\begin{figure}
\includegraphics[width = 8.6 cm]{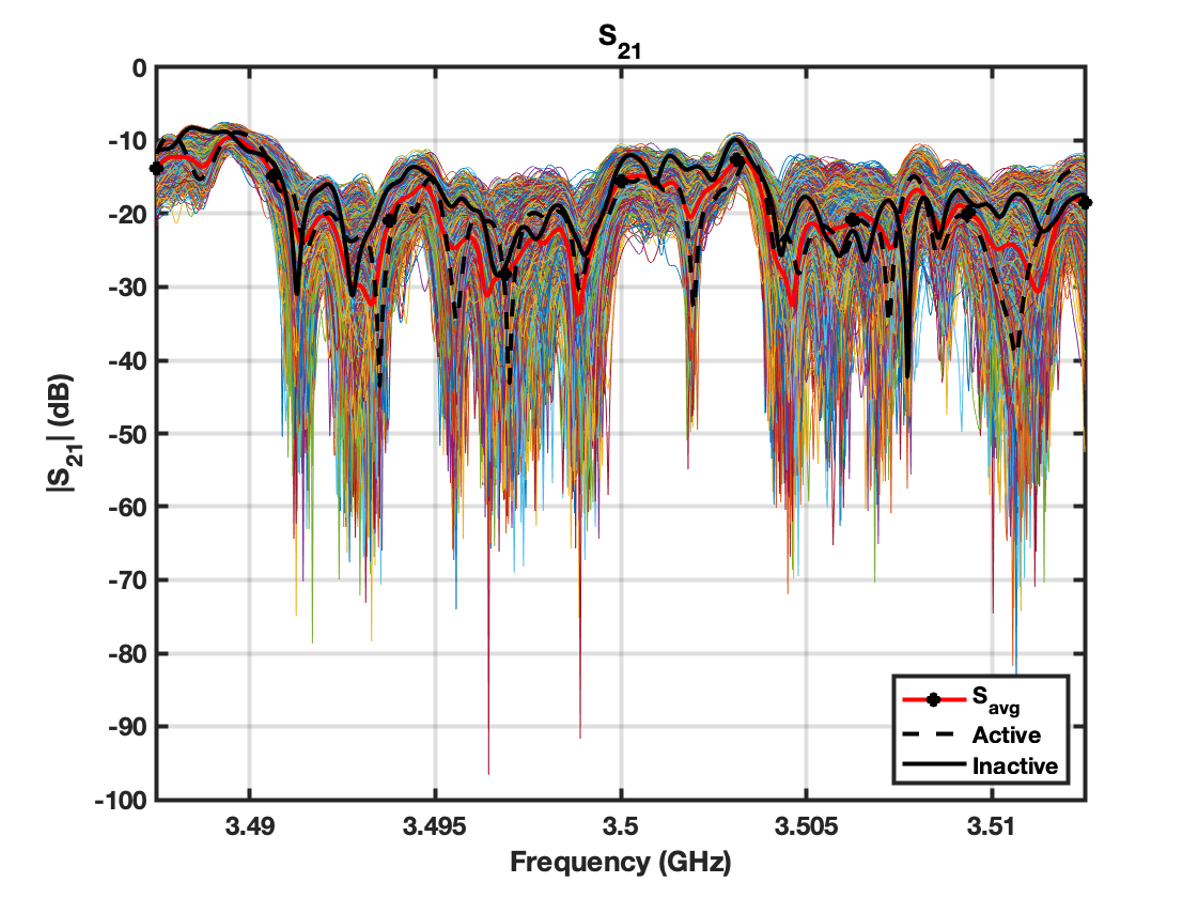}
\caption{\label{fig:distributions} \bf Ensemble of 4000 $|S_{21}|$ realizations from a combination of doubly random power spectrum and biased coin toss approaches. \normalfont The mean value is shown as the solid red line, the case with all elements active is shown as the solid black line and the case with all elements inactive is shown as the dashed black line.}
\end{figure}

A small exponent will excite lower spatial frequencies, while a larger exponent will excite higher spatial frequencies. the number of active elements was allowed to change with each trial. We generally used a combination, with half the ensemble generated through an inverse power spectrum and the other half generated through a random biased coin toss. The ensemble for a set of 4000 realizations is shown in Figure \ref{fig:distributions}. The doubly random methods allow the number of active elements to change and provide a wider range of responses than the deterministic methods. This can be seen as the distribution from the biased coin toss covers the entire area between the inactive (all 0s) and active (all 1s).

\section{Power Optimization Results}
The optimization algorithm was successfully performed with different cavity configurations and different performance metrics. From Figure \ref{fig:distributions}, we can see that the bandwidth of the narrowest null is $\sim$ 500 kHz and the closest spacing between nulls is $\sim$ 1 MHz. This matches the observed trends over the full 1 GHz measurement window, with typical bandwidths of 0.5-1 MHz and spacings of 1-2 MHz. The metric was chosen to be the average power measured at port 2, $P_2$, when driving an input from some combination of ports 1 and 3. To isolate narrow band features, our initial metric bandwidth was chosen to be 500 kHz.

\begin{figure*}
\includegraphics[width = 17.2cm]{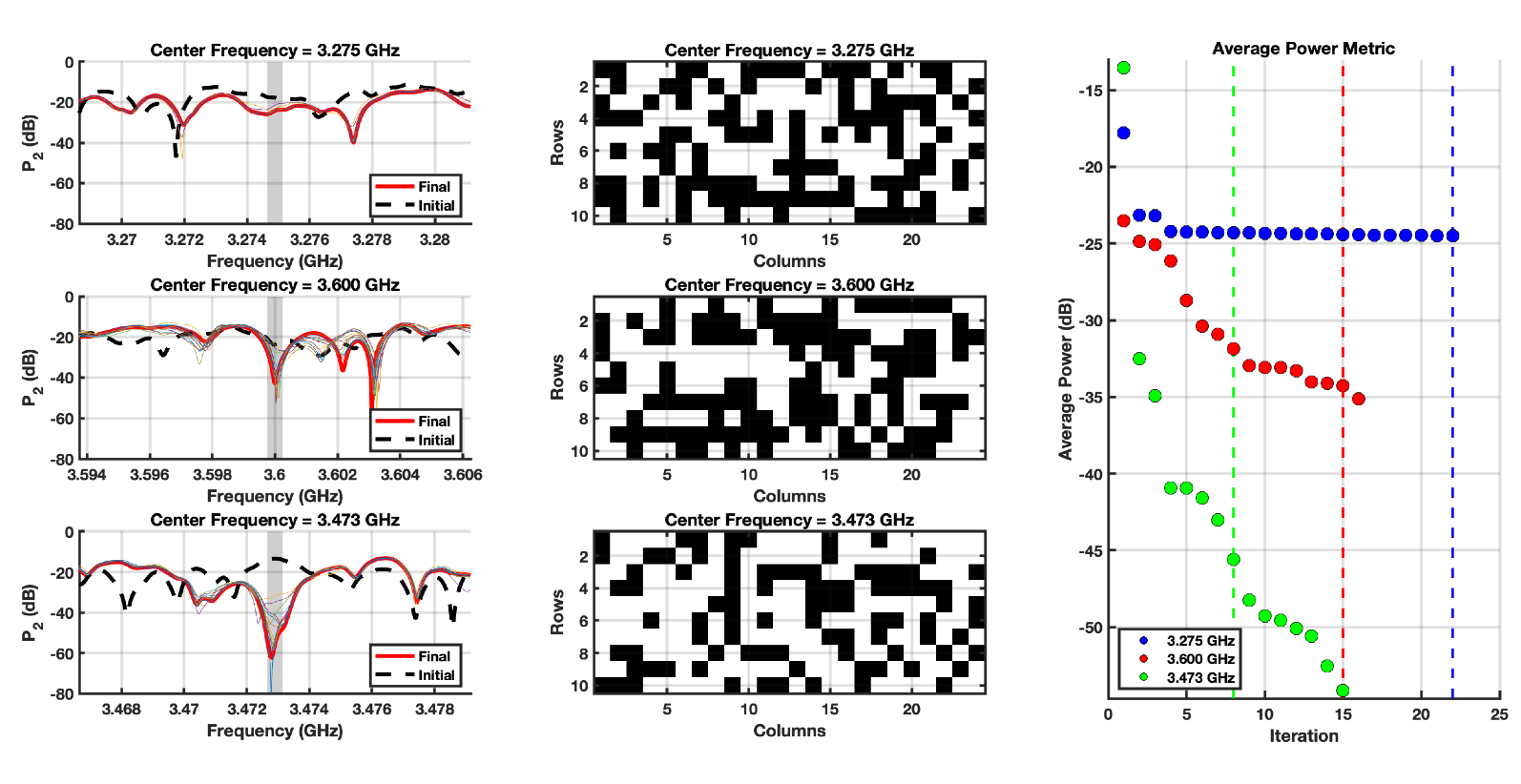}
\caption{\label{fig:singleFreqPort1} \bf Optimization results when driving port 1 with the metric defined over selected frequency bands. \normalfont A 500 kHz bandwidth was used and the center frequencies are 3.275 GHz, 3.600 GHz, and 3.473 GHz. The measured $P_2$ responses are given in the left hand side and show the initial all 0s state as the dashed black line, the final state as the bold red line and the incremental states as the remaining lines. The region where the metric was evaluated is shown as the shaded gray region. The optimal set of commands is shown in the middle and the evolution of the metric is shown in the right hand side. The worst case improvement was 7 dB and the best case improvement was 40 dB.}
\end{figure*}

For initial validation of the algorithm, we used single frequency bands. To ensure that the band selection was not biased to only bands where the metasurface performed well, the center frequency for the metric was chosen as a random draw during each experimental run. Figure \ref{fig:singleFreqPort1} shows the results for optimizing at 3 different frequencies when driving port 1. This figure shows that we were able to realize deep nulls in many arbitrary locations and were able to reduce the power by at least 4 dB in each band attempted.

\begin{figure*}
\includegraphics[width = 17.2cm]{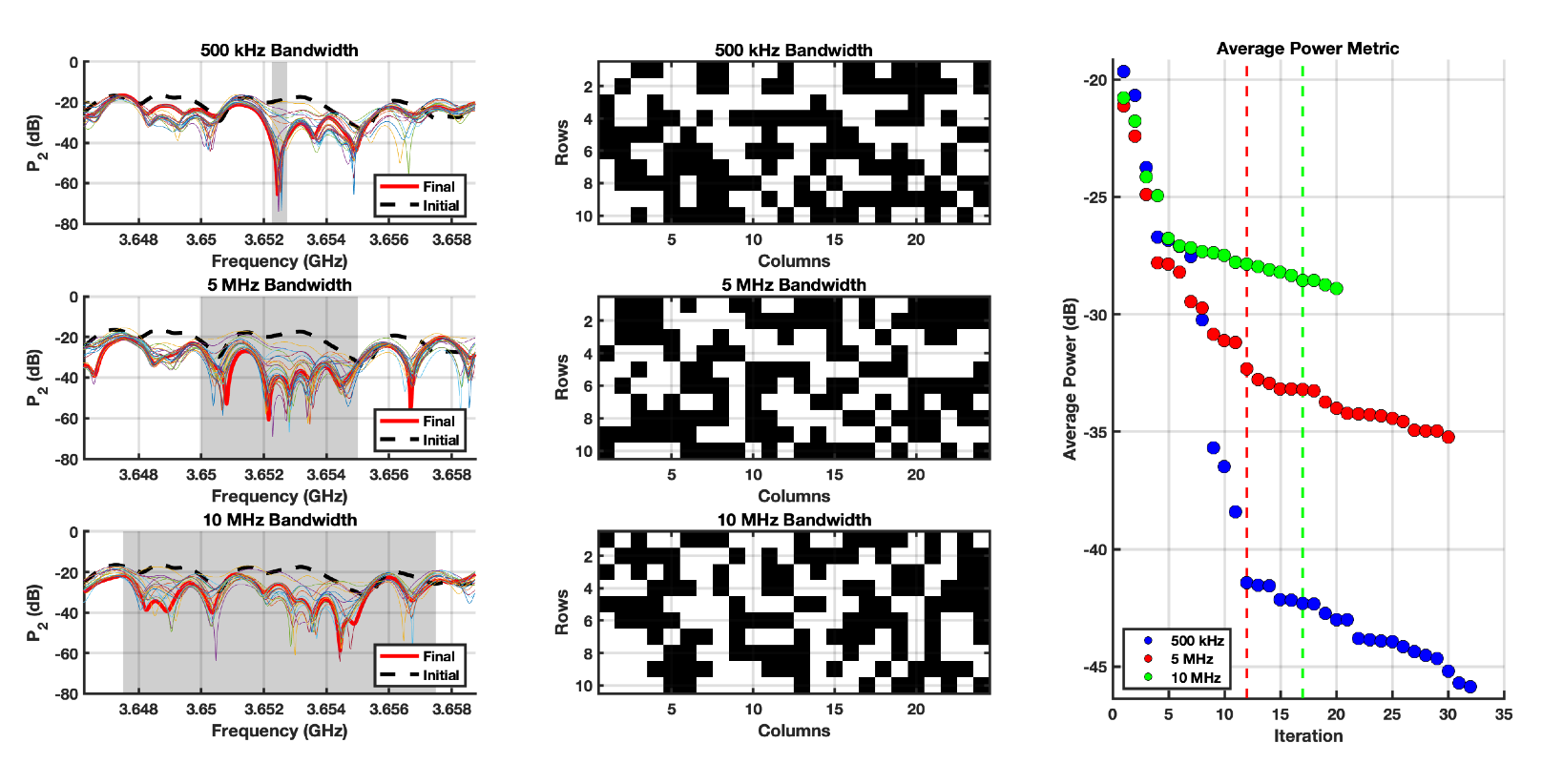}
\caption{\label{fig:singleFreqMuliBWPort3} \bf Optimization results when driving port 3 with with the metric defined over different  bandwidths. \normalfont  The metric was centered at 3.652 GHz and bandwidths of 500 kHz, 5 MHz, and 10 MHz were used. The measured $P_2$ responses are given in the left hand side and show the initial all 0s state as the dashed black line, the final state as the bold red line and the incremental states as the remaining lines. The region where the metric was evaluated is shown as the shaded gray region. The optimal set of commands is shown in the middle and the evolution of the metric is shown in the right hand side. The worst case improvement was 6 dB and the best case improvement was 40 dB.}
\end{figure*}

The next step was to determine how the optimization algorithm responded when the metric was defined with different bandwidths. Figure \ref{fig:singleFreqMuliBWPort3} shows the optimization results with 500 kHz, 5 MHz, and 10 MHz bandwidths centered at 3.652 GHz when driving port 3. This figure shows that suppression of $P_2$ decreases as the bandwidth increases. This is not surprising as a wider bandwidth means more modes and therefore more Degrees Of Freedom (DOF) that need to be matched. We can also see how the shape of the spectrum changes as features outside the narrow bandwidth metric get included in the wider bandwidth metrics.

\begin{figure*}
\includegraphics[width = 17.2cm]{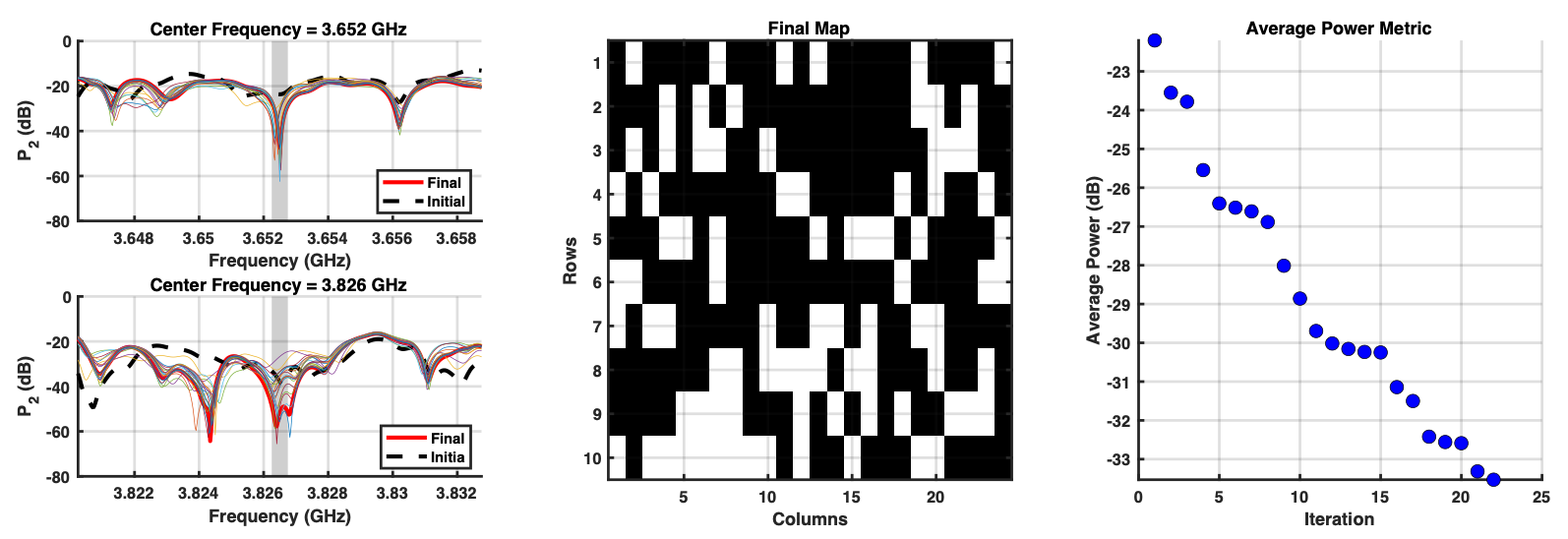}
\caption{\label{fig:dualband1} \bf Optimization results when driving port 1 with the metric defined over 2 separated frequency bands. \normalfont The bandwidths were 500 kHz and the center frequencies were 3.652 GHz and 3.826 GHz. The measured $P_2$ responses are given in the left hand side and show the initial all 0s state as the dashed black line, the final state as the bold red line and the incremental states as the remaining lines. The region where the metric was evaluated is shown as the shaded gray region. The optimal set of commands is shown in the middle and the evolution of the metric is shown in the right hand side.  The total reduction in average power was 10 dB.}
\end{figure*}

Next, we evaluated the performance when the metric was defined over 2 separated frequency bands. Figure \ref{fig:dualband1} shows the optimization results when driving port 1 and defining the metric as the average power contained in two separated frequency bands. The suppression capability when using multiple frequency bans is reduced; as in the case with a wider bandwidth metric, we are including more features and need to match more DOF.

\begin{figure*}
\includegraphics[width = 17.2cm]{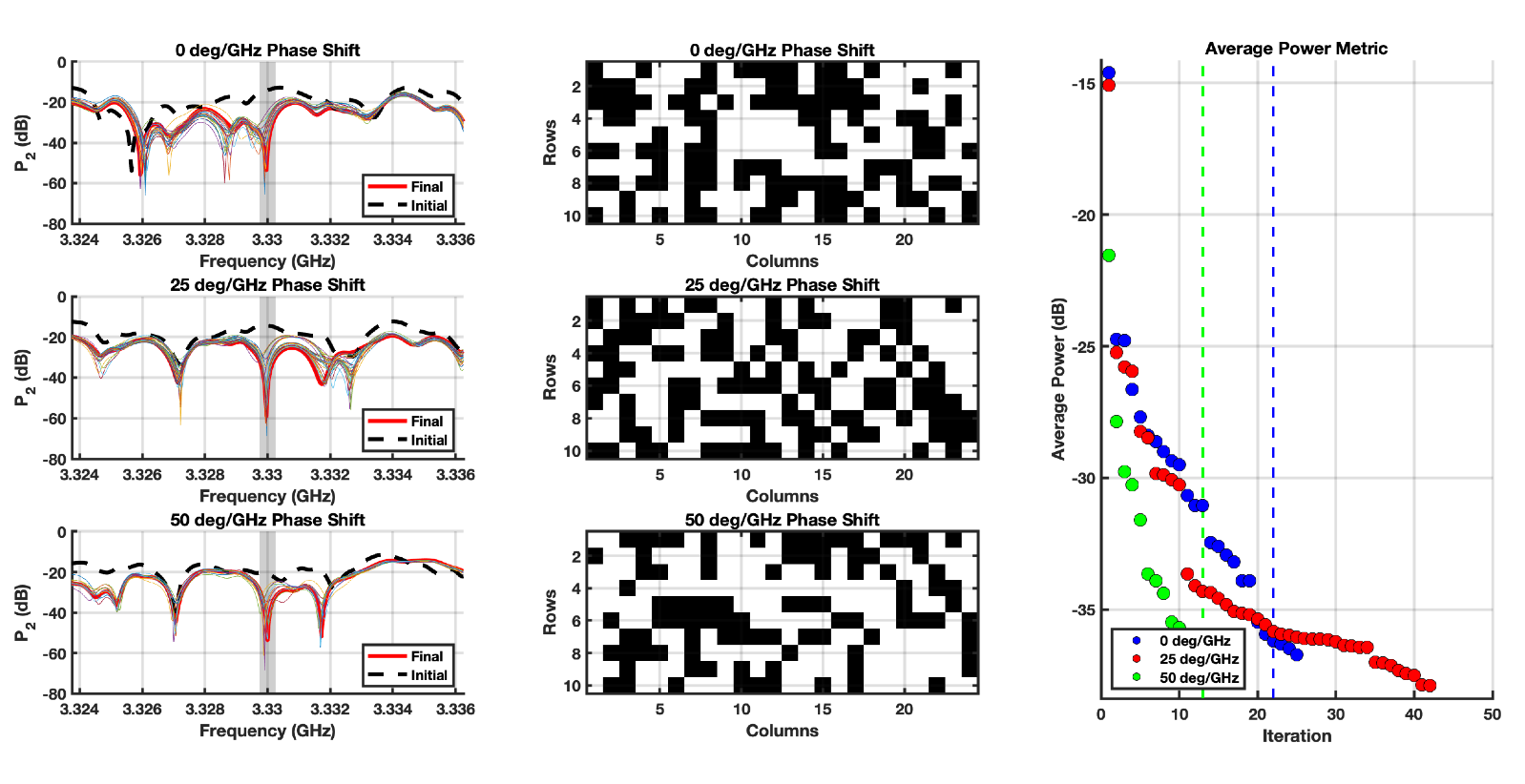}
\caption{\label{fig:phaseshift} \bf Optimization results when driving ports 1 and 3 collectively with a relative phase shift. \normalfont The metric was defined over a 500 kHz bandwidth centered at 3.33 GHz. The measured $P_2$ responses are given in the left hand side and show the initial all 0s state as the dashed black line, the final state as the bold red line and the incremental states as the remaining lines. The region where the metric was evaluated is shown as the shaded gray region. The optimal set of commands is shown in the middle and the evolution of the metric is shown in the right hand side. The impact of the phase shift can be seen in the difference of the initial $P_2$ response between the various cases and the total power reduction was $\sim$ 20 dB in each case.}
\end{figure*}

The last step for the iterative algorithm was to observe the performance when ports 1 and 3 were driven simultaneously with a relative phase shift. Figure \ref{fig:phaseshift} shows the results with phase shifts of 0 deg/GHz, 25 deg/GHz, and 50 deg/GHz. This shows we are able to generate coldspots in complex cases where the inputs themselves add spatial and angular diversity.

\begin{figure*}
\includegraphics[width = 17.2cm]{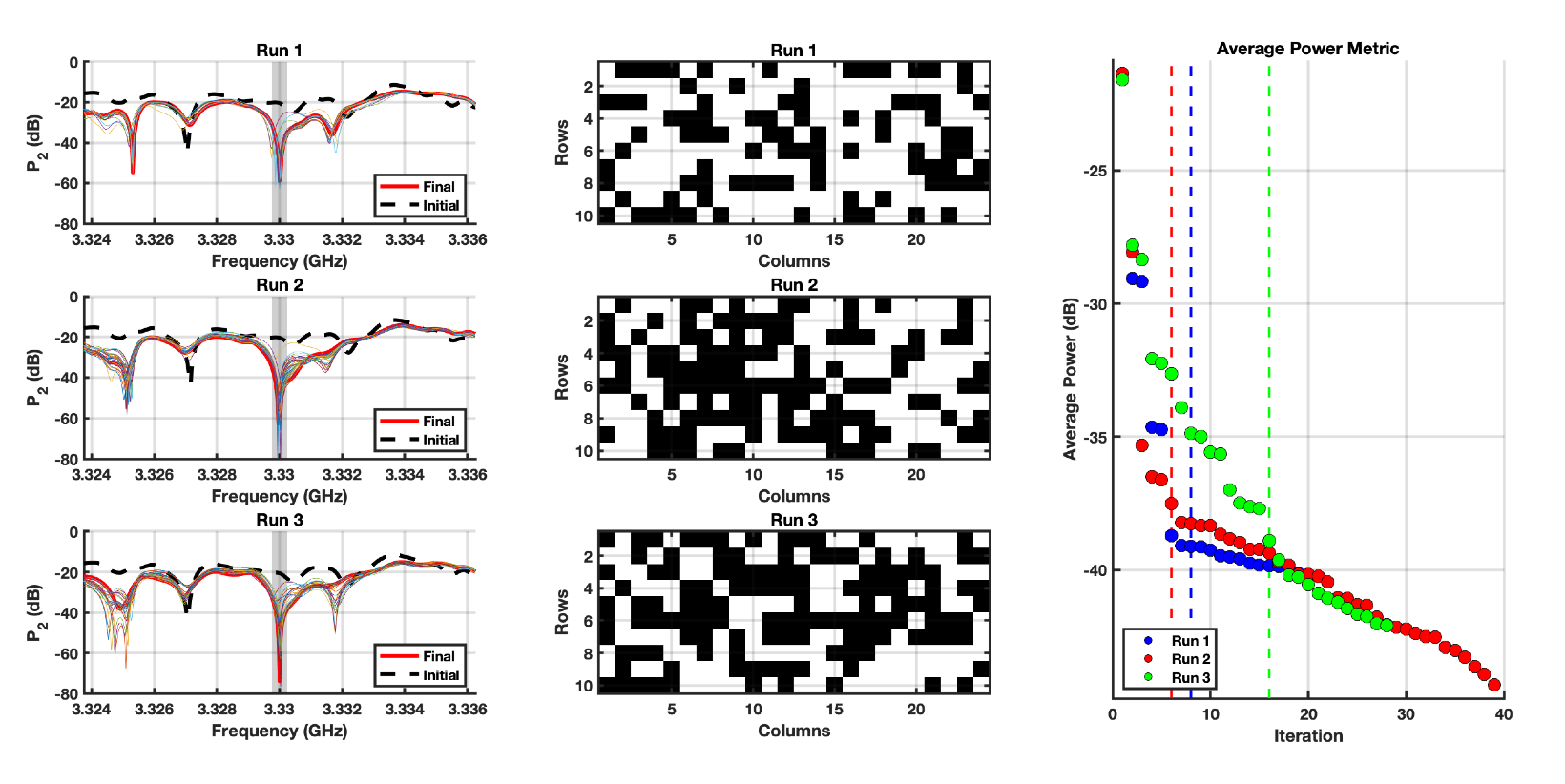}
\caption{\label{fig:phaseshift_repeat} \bf Results of repeating an optimization experiment. \normalfont Optimization results when driving ports 1 and 3 collectively with a 50 deg/GHz relative phase shift with the metric defined over a 500 kHz bandwidth centered at 3.33 GHz. The measured $P_2$ responses are given in the left hand side and show the initial all 0s state as the dashed black line, the final state as the bold red line and the incremental states as the remaining lines. The region where the metric was evaluated is shown as the shaded gray region. The optimal set of commands is shown in the middle and the evolution of the metric is shown in the right hand side. This shows the results trended in the same direction, the final achieved state depends on }
\end{figure*}

Finally, we look at the repeatability of the optimization process and determine if we can recover the same final state each time. Figure \ref{fig:phaseshift_repeat} shows the results from repeating an optimization when driving ports 1 and 3 collectively with a 50 deg/GHz phase shift. The final states are different; however, the metrics followed the same path and the differences are due to the algorithm declaring convergence and stopping at different times. Increasing the convergence criteria would likely result in more closely related final states.

\section{Coherent Perfect Absorption State Generation and Verification}
\begin{figure}
\includegraphics[width = 8.6cm]{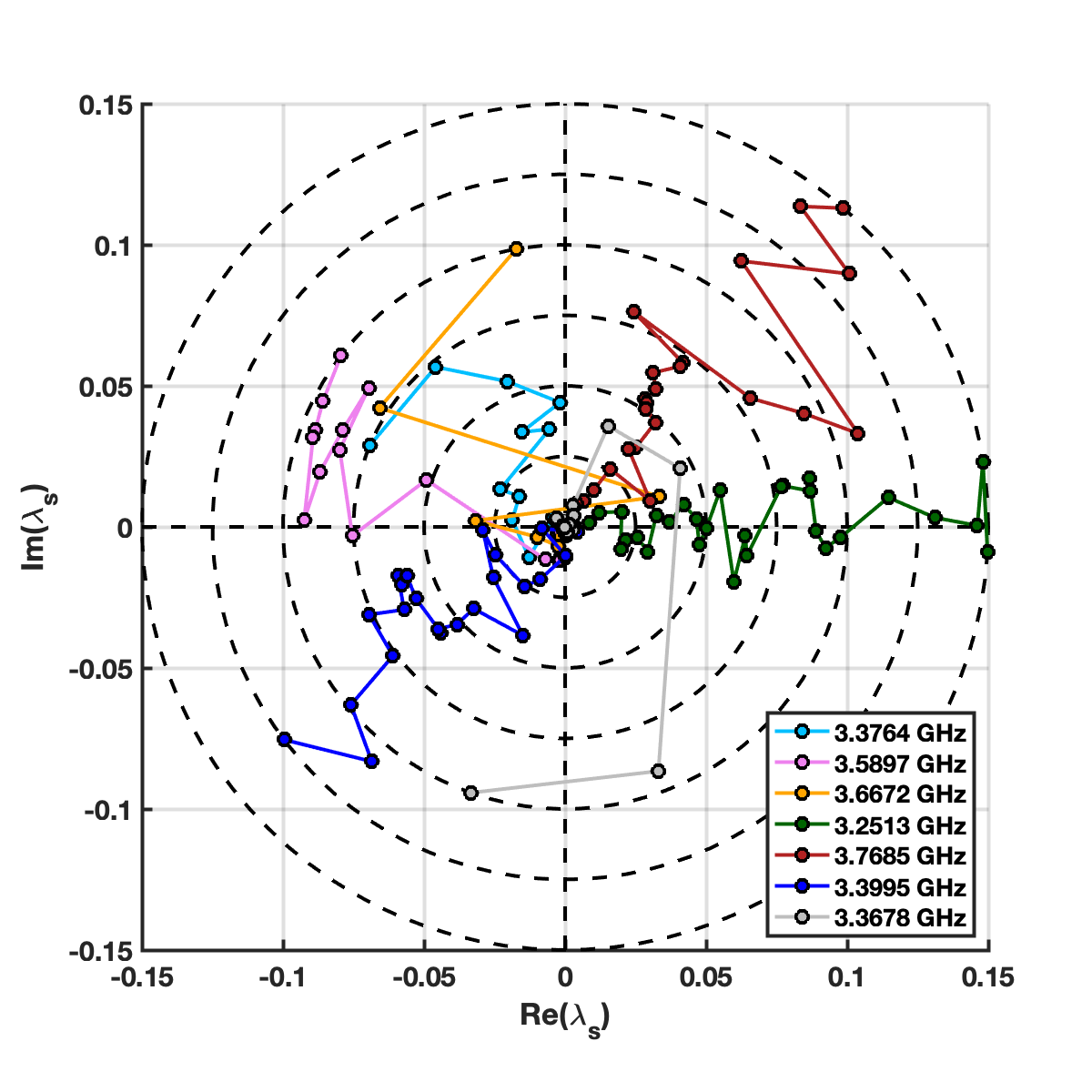}
\caption{\label{fig:eig_directivity} \bf Directed eigenvalue trajectories. \normalfont Each curve represents an attempt to drive the eigenvalues towards the origin.} 
\end{figure}

A Coherent Perfect Absorption (CPA) state is not guaranteed at any specific frequency, so the approach needs to identify candidates. We repeated the iterative optimization algorithm from the coldspot generation but initialized it by applying a random command set to the metasurface and then finding the eigenvalue with magnitude closest to a pre-selected value. Figure \ref{fig:eig_directivity} shows the results for 7 experiments with the eigenvalue initialized so that $0.075 \leq |\lambda_s| \leq 0.15$. This figure clearly shows the directed random walk nature of the iterative algorithm and demonstrates that the eigenvalues can in general move significantly.

Verifying the CPA state requires exciting the scattering system with the corresponding eigenvalue. We used an Agilent N5242A PNA-X Network Analyzer configured for 2-independent source operation. This configuration requires selecting the appropriate signal path on the setup menu and making the jumper connections on the back of the instrument as per the network analyzer documentation. An external phase shifter was connected between port 1 and the cavity. The CPA state condition was tuned by adjusting the relative amplitude between the elements with the power control on the network analyzer and the relative phase between the elements with the external phase shifter. We can perform parameter sweeps to determine the sensitivity of the CPA to frequency, relative phase, relative amplitude, and metasurface commands.

Because there are 2 independent sources, $S$-parameter measurements are not available in this configuration and we need to make use of the network analyzer receivers. The network analyzer has reference and test port receivers to measure incoming and outgoing signals. $R_1$ measures the reference signal out of port 1 and $R_2$ measures the reference signal out of port 2. $A$ measures the signal into port 1 and $B$ measures the signal into port 2. The metric we are interested in is the power ratio of the total outgoing power from the cavity to the total incoming power from the network analyzer,  $P_{\text{out}}/P_{\text{in}}$ = $(A + B)/(R_1 + R_2)$.


%